\documentclass[final,5p,times,twocolumn,sort&compress]{elsarticle}
\usepackage{graphicx}
\usepackage{subfigure}
\usepackage[usenames]{color}
\usepackage{simplemargins}
\usepackage{amssymb}
\usepackage{amsmath}
\usepackage{amsfonts}
\usepackage{multicol}
\usepackage{url}

\journal{Journal of Computational Science}

\begin{document}
\begin{frontmatter}

 \title{Twitter reciprocal reply networks exhibit assortativity with respect to happiness } 
 
 \author{Catherine A. Bliss}
 \ead{Catherine.Bliss@uvm.edu}
  \author{Isabel M. Kloumann}
  \ead{Isabel.Kloumann@uvm.edu}
     \author{Kameron Decker Harris }
     \ead{Kameron.Harris@uvm.edu}
     \author{Christopher M. Danforth}
     \ead{Chris.Danforth@uvm.edu}
    \author{Peter Sheridan Dodds}
 \ead{Peter.Dodds@uvm.edu}
 \address{Department of Mathematics and Statistics, Vermont Complex Systems Center\\ \& the Vermont Advanced Computing Core, 
 University of Vermont, Burlington, VT, 05405}

\begin{abstract}
\noindent The advent of social media has provided an extraordinary, if imperfect, 'big data' window into the form and evolution of social networks. Based on nearly 40 million message pairs posted to Twitter between September 2008 and February 2009, we construct and examine the revealed social network structure and dynamics over the time scales of days, weeks, and months. At the level of user behavior, we employ our recently developed hedonometric analysis methods to investigate patterns of sentiment expression. We find users' average happiness scores to be positively and significantly correlated with those of users one, two, and three links away. We strengthen our analysis by proposing and using a null model to test the effect of network topology on the assortativity of happiness. We also find evidence that more well connected users write happier status updates, with a transition occurring around Dunbar's number. More generally, our work provides evidence of a social sub-network structure within Twitter and raises several methodological points of interest with regard to social network reconstructions.\\
\end{abstract}

\end{frontmatter}

\section{Introduction}
\vspace{-3mm}
Social network analysis has a long history in both theoretical and
applied settings~\cite{wasserman1994}.
During the last 15 years,
and driven by the increased availability of real-time, in-situ data 
reflecting people's social interactions and choices, 
there has been an explosion of research activity around social phenomena,
and many new techniques for characterizing large-scale social networks have emerged.  
Numerous studies have examined the structure of online social networks
in particular, such as blogs, Facebook, and
Twitter~\cite{Gjoka2010, Viswanath2009, Papacharissi2009, Dodds2010,
Java2009, bakshy2011everyones, Bollen2011, Bollen2011a, cha2010,
Dodds2011, Guo_2009, huberman2008, kim2009, Kwak2010, thelwel2010,
weng2010, Lee2011, Kleinberg2012}. 

In a series of analyses of the Framingham Heart Study data and the
National Longitudinal Study of Adolescent Health, Christakis, Fowler,
and others have examined how qualities such as happiness, obesity,
disease, and habits (e.g., smoking) are correlated within social
network neighborhoods~\cite{christakis2007, Fowler2008,
Christakis2008, Rosenquist2010, Hill2010, Christakis2011}. 
The authors' additional assertion of contagion, however, has been
criticized primarily on the basis of the difficulties to be found in distinguishing
these phenomena from homophily~\cite{noel2011, lyons2011,
Shalizi2011}. The observation that social networks
exhibit assortativity with respect to these traits evidently requires further study and leads us to explore potential mechanisms. Advances would naturally 
provide further insight into the nature of how social groups 
influence individual behavior and vice versa.

Our focus in the present work is the social
network of Twitter users. With the abundance of available data,
Twitter serves as a living laboratory for studying contagion and
homophily~\cite{romero2011}. As a requisite step towards these goals,
we first define sub-networks of Twitter users suitable to such study
and, second, examine whether assortativity is observed in these
sub-networks. Before describing our methods, we provide a brief
overview of Twitter, related work, and the challenges associated with
social network analysis in this arena.

Twitter is an online, interactive social media platform
in which users post tweets, micro-blogs with a 140 character limit. 
Since its inception in 2006, 
Twitter has grown to encompass over 200 million accounts, with over 100
million of these accounts currently active as of October 2011,
and with some users having garnered over 10 million
followers \cite{blog_twitter}. 
Tweets are open online by default, 
and are also broadcast directly to a user's followers.
Users may express interest in a tweet by
retweeting the message to their followers. 
Alternatively, followers
may reply directly to the author. 

Understanding the topology of the Twitter network, the
manner in which users interact and the diffusion of information
through this media is challenging, both computationally and
theoretically. One of the central issues in characterizing the topology of any
network representation of Twitter lies in defining the criteria for
establishing a link between two users. The majority of previous
studies have examined the topology of and information cascades on the
Twitter follower network \cite{bakshy2011everyones, cha2010,
Kwak2010}, as well as on networks derived from mutual following
 \cite{Bollen2011}.
However, the follower network is not the only
representation of Twitter's social network, and its structure can be
misleading  \cite{Goncalves2011}. For example, in a study of over 6 million users, Cha et
al. \cite{cha2010} found that users with the highest follower counts
were not the users whose messages were most frequently retweeted. This
suggests that such popular users (as measured by follower count) may
not be the most influential in terms of spreading information, and this
calls into question the extent to which users are influenced by those
that they follow \cite{watts2007a}. Of further concern is the finding of low reciprocity
within follower networks. Kwak et al.\ found very few individuals who
followed their followers \cite{Kwak2010}. As a result, trying to infer meaningful
influence and contagion in such a network is difficult. 

\begin{figure}[htp]{
\centering 
\subfigure[Followers]{\includegraphics[width=.2\textwidth]{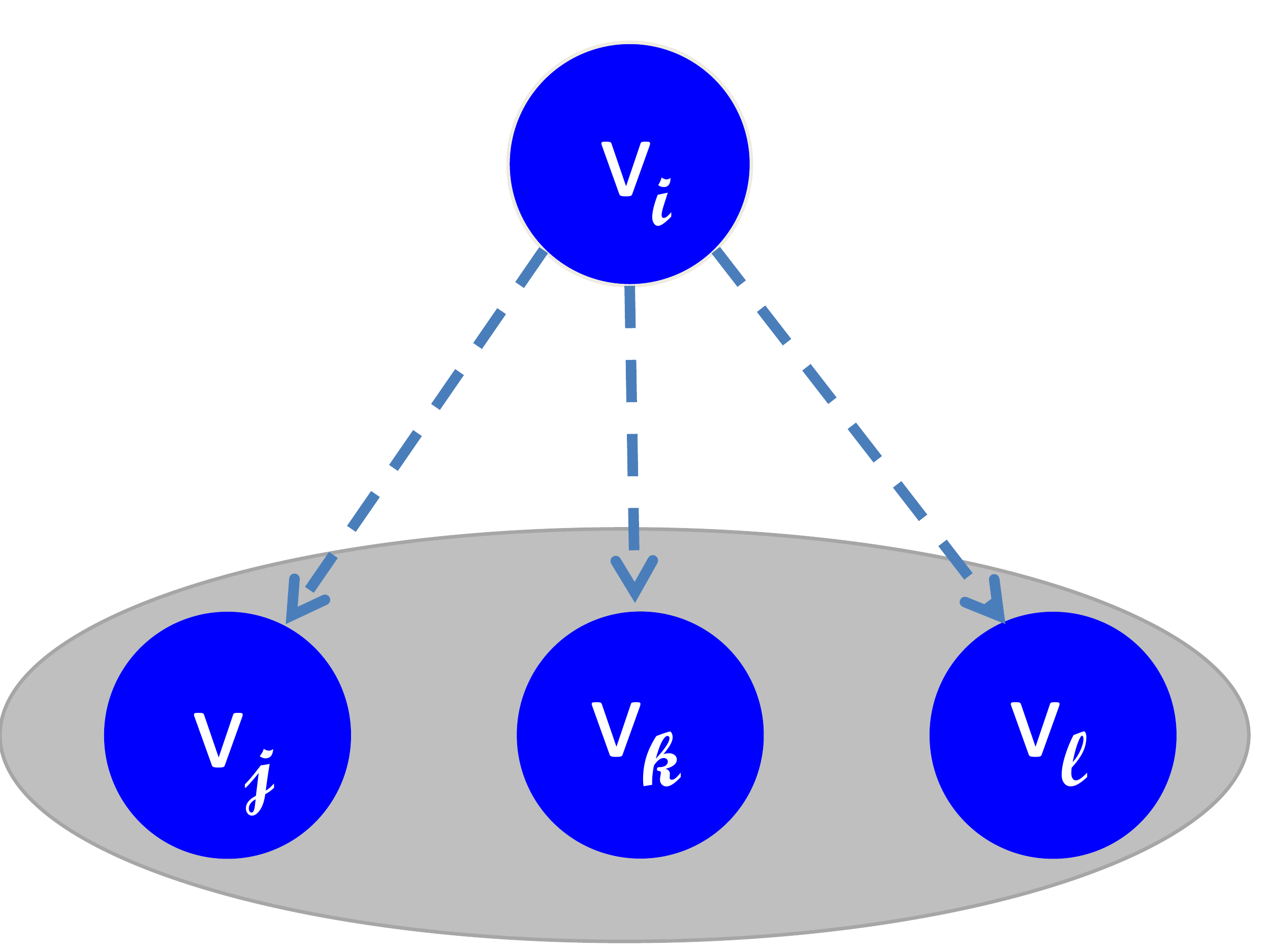}}
\subfigure[Interaction]{\includegraphics[width=.2\textwidth]{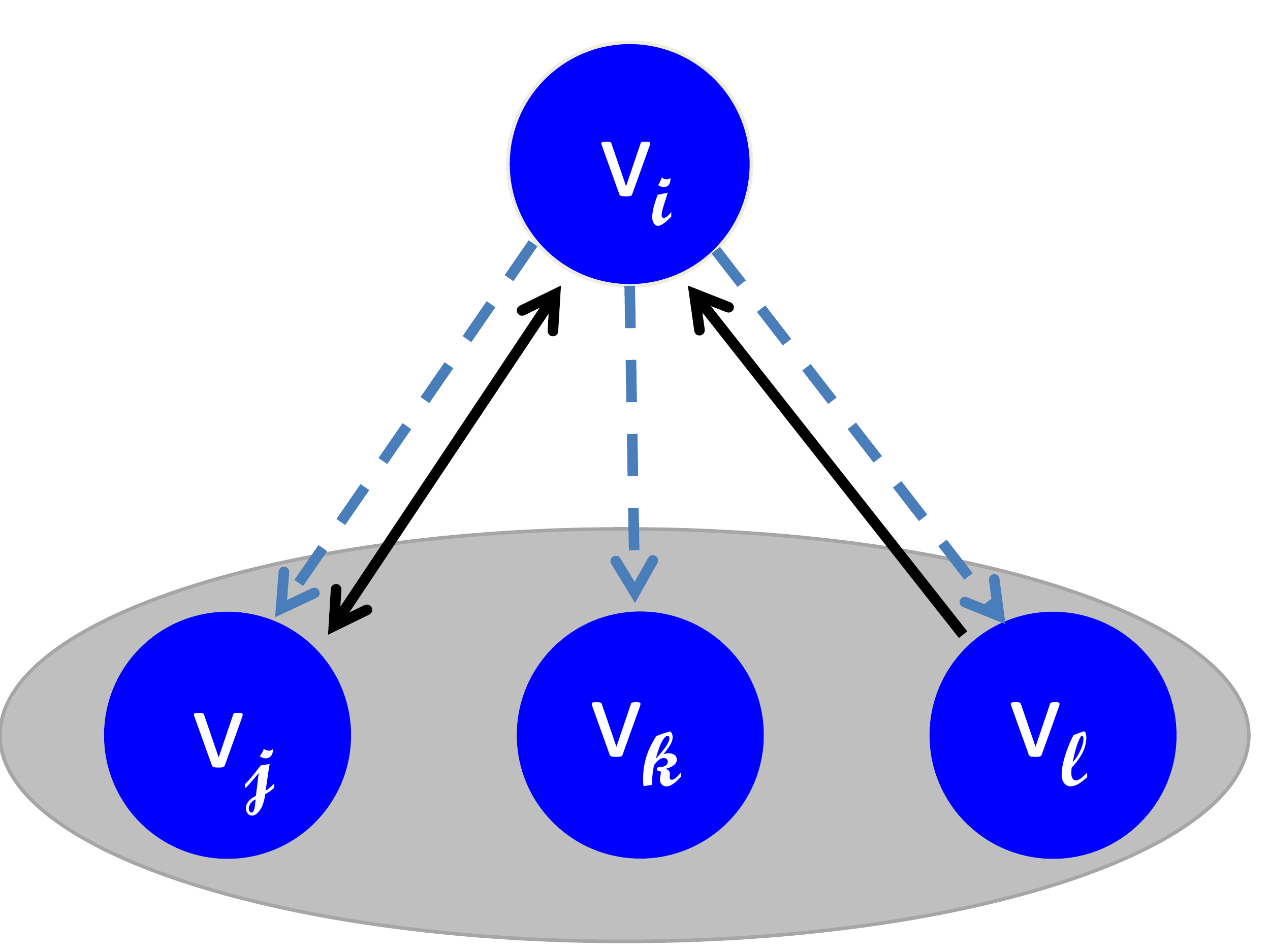}}
\caption{(a) Follower network: The follower network is generated by declared following choices, absent any messages being sent. If user $v_i$ broadcasts tweets to followers $v_j, v_k$ and $v_{\ell}$ (represented by the dashed, blue arrow) $v_i$ would be connected to each of $v_j, v_k$ and $v_{\ell}$ by a directed link in a follower network. (b) Reciprocal-reply network: Directed replies are represented by a solid black arrow. When considering the interaction between users, a reply (i.e., $v_{\ell}$ replies to $v_i$) provides evidence of a directional interaction between nodes. We mandate a stronger condition for interaction, namely reciprocal replies (i.e., $v_j$ replies to $v_i$ and vice versa) over a given time period. Thus $v_i$ and $v_j$ are connected in the reciprocal reply network that we construct.}
\label{fig:follower_bid_network}}
     \end{figure}

While popular users and their many followers clearly exhibit an
affiliation, they do not necessarily interact, as there are different relationships implicated by broadcasting (tweeting), sending a message
(@someone), and replying to a message. As an example, we consider a
user represented by node $v_i$ which has three followers, represented
by $v_j, v_k,$ and $v_{\ell}$ as shown in Fig. \ref{fig:follower_bid_network}a. 
When a user broadcasts tweets to
their many followers, as represented by the directed arrow in Fig. \ref{fig:follower_bid_network}a, 
this does not imply that followers read or respond to these
tweets.  Followers $v_j,v_k,$ and $v_{\ell}$ receive all tweets
broadcast by node $v_i$, but this provides no guarantee of
interaction. Suppose, though, that we observe that $v_{\ell}$ replies
to $v_i$ as shown in Figure \ref{fig:follower_bid_network}b.  This
provides evidence (but not proof) that the user represented by $v_{\ell}$ has indeed
received a tweet from $v_i$ and is sufficiently motivated to create a
response to $v_i$. Although a directional network based on these
replies can be created, such a directional interaction, however, does
not suggest reciprocity between the nodes. In this example, we have no
evidence that $v_i$ has, in any way, considered or even read such a
response from his/her follower.

We conclude that following and unreciprocated replies are not
sufficient for interaction and present an alternative means by which
to derive a social network from Twitter messages, via reciprocal
replies. In our reciprocal-reply network, two nodes, $v_i$ and $v_j$,
are connected if $v_i$ has replied to $v_j$ and $v_j$ has replied to
$v_i$ at least once within a given time period of consideration. In Figure
\ref{fig:follower_bid_network}b, the nodes $v_i$ and $v_j$ meet this
criterion. 
 
Another challenge in characterizing the topology of any network representation of Twitter concerns determining how long a link between two users in the network should persist. Including stale user-user interactions in the network mistakenly creates an inaccurate portrayal of the current state of the system; this is typically referred to as the ``unfriending problem'' \cite{noel2011}. Not only will network statistics such as the number of nodes, average degree, maximum degree and proportion of nodes in the giant component be artificially inflated due to superfluous, no-longer-active links  \cite{noel2011, Grannis2010}, but the degree distribution will also be distorted. Kwak et al.\ \cite{Kwak2010} found that the degree distribution for a Twitter follower network deviated from a power law distribution due to an overabundance of high degree nodes resulting from an accumulation of ``dead-weight'' in the network.   

Additional problems are encountered if one uses accumulated network data to measure assortativity with respect to a trait (e.g., happiness). As an example, consider a network in which two users are connected because they interacted during the last week of a year-long study. Including this user-user pair in the list of pairs to compute assortativity for the entire network blurs the relationship between more consistent and repeated interactions that occurred throughout the timespan of the study. Further complications arise when averaging a user's trait over a large time scale (i.e., averaging happiness over a 6 month or 12 month timespan). Detecting changes in users' traits over time and how these may (or may not) be correlated with nearest neighbors' traits is of fundamental importance; accumulated network data occludes exactly the interactions we are looking to understand. Recognizing that, due to practical limitations, accumulation of network data must occur on some scale, we analyze users in day, week, and month reciprocal reply networks. By examining networks constructed at smaller time scales and calculating users' happiness scores based on tweets made only during that time period, we aim to take a more dynamic view of the network. 

In addition to defining reciprocal reply networks and advocating for their use, we also seek to describe how happiness is distributed in the reciprocal reply networks of Twitter. Previous hedonometric work with Twitter data has revealed cyclical fluctuations in average happiness at the level of days and weeks, as well as spikes and troughs over a time scale of years corresponding to events such as U.S. Presidential Elections, the Japanese tsunami and major holidays \cite{Dodds2011, golder2011, miller2011}. Other studies have examined changes in valence of tweets associated with the death of Michael Jackson \cite{kim2009}, changes in the U.S. Stock Market \cite{Bollen2011a}, the Chilean Earthquake of 2010, and the Oscars \cite{thelwel2010}. In the present work, we seek to understand localized patterns of happiness in the Twitter users' social network.

Understanding how emotions are distributed through social networks, as well as how they may spread, provides insight into the role of the social environment on individual emotional states of being, a fundamental characteristic of any sociotechnical system. Bollen et al.\ \cite{Bollen2011} examine a reciprocal-follower network using Twitter and suggest that Subjective Well-Being (SWB), a proxy for happiness, is assortative. Building on their work, we address whether happiness is assortative in reciprocal-reply networks. We also test the hypothesis of Christakis and Fowler  \cite{Christakis2011} who find evidence that the assortativity of happiness may be detected up to three links away. In doing so, we raise an additional point which is not specific to Twitter networks, but rather relates to empirical measures of assortativity in general. Relatively few studies have employed a null model for calculating the pairwise correlations (e.g., happiness-happiness). We devise a null model which maintains the topology of the network and randomly permutes happiness scores attached to each node. By randomly permuting users' happiness scores, we can detect what effect, if any, network structure has on the pairwise correlation coefficient.

We organize our paper as follows: In Section 2, we describe our data set, the algorithm for constructing reciprocal-reply networks, network statistics used for characterizing the networks, and our measure for happiness. We propose an alternative means by which to detect social structure and argue that our method detects a large social sub-network on Twitter. In Section 3, we describe the structure of this network, the extent to which it is assortative with respect to happiness and the results of testing assortativity against a null model. In Section 4, we discuss these findings and propose further investigations of interest.

\vspace{-3mm}
\section{Methods}
\vspace{-3mm}
\subsection{Data}
\begin{figure}[htp!]{
\centering     
\includegraphics[width=.45\textwidth]{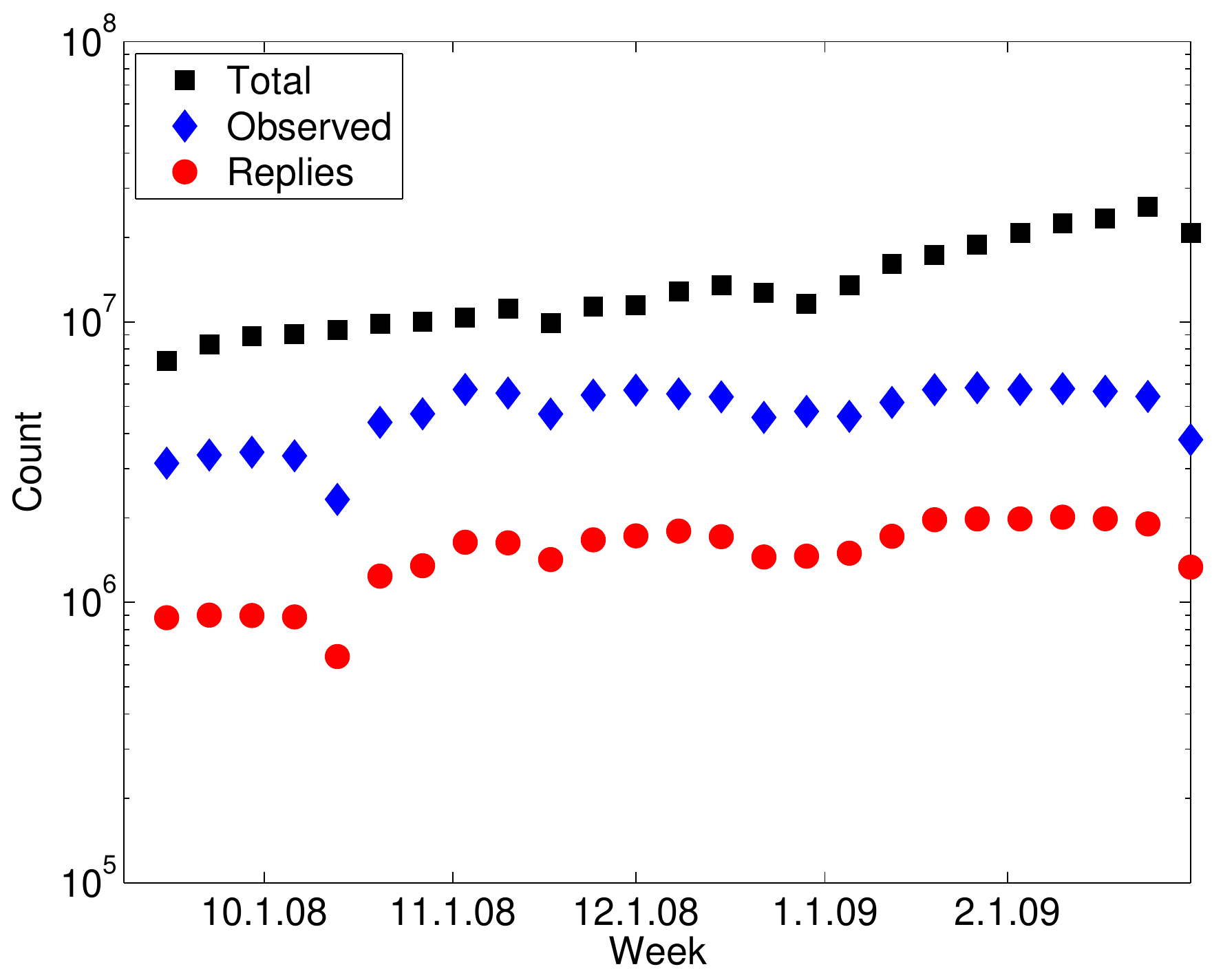}
\caption{Tweet counts are plotted for the weeks between September 2008 and February 2009. The three curves represent the total, those that we observed and the number of the observed tweets that constituted replies. }
\label{fig:data_set}}
     \end{figure}

From September 2008 to February 2009, we retrieved over 100 million tweets from the Twitter streaming API service.\footnote{Data was received in XML format.} While the volume of our feed from the Twitter API increased during this study period, the total number of tweets grew at a faster rate (Fig. \ref{fig:data_set}). During this time period, we estimate that we collected roughly 38\% of all tweets.\footnote{We calculated the total number of messages as the difference between the last message id and the first message id that we observe for a given week. This provides a reasonable estimate of the number of tweets made per week, as message ids were assigned (by Twitter) sequentially during the time period of this study.} The number of messages and percent of which were replies are reported in Table A\ref{table:data set}. For the remainder of this paper, we restrict our attention to the nearly 40 million message-reply pairs within this data set and the users who authored these tweets.

The data received from the Twitter API service for each tweet contained separate fields for the identification number of the message (message id), the identification number of the user who authored the tweet (user id), the 140 character tweet, and several other geo-spatial and user-specific metadata. If the tweet was made using Twitter's built-in reply function,\footnote{Twitter has a built-in reply function with which users reply to specific messages from other users. Tweets constructed using Twitter's reply function begin with `@username', where `username' is the Twitter handle of the user being replied to; the user and message ids of the tweet being replied to are included in the reply message's metadata from the Twitter API. Users often informally reply to or direct messages to other users by including said users' Twitter handles in their tweets. In such cases, however, no identification information about the ``mentioned'' user is included in the API parameters for these tweets (only their Twitter handle is) and we exclude such exchanges when building the reciprocal reply network.} the identification number of the message being replied to (original message id) and the identification of the user being replied to (original user id) were also reported. 

\begin{figure}[htp!]
\centering     
\subfigure{\includegraphics[width=.2\textwidth]{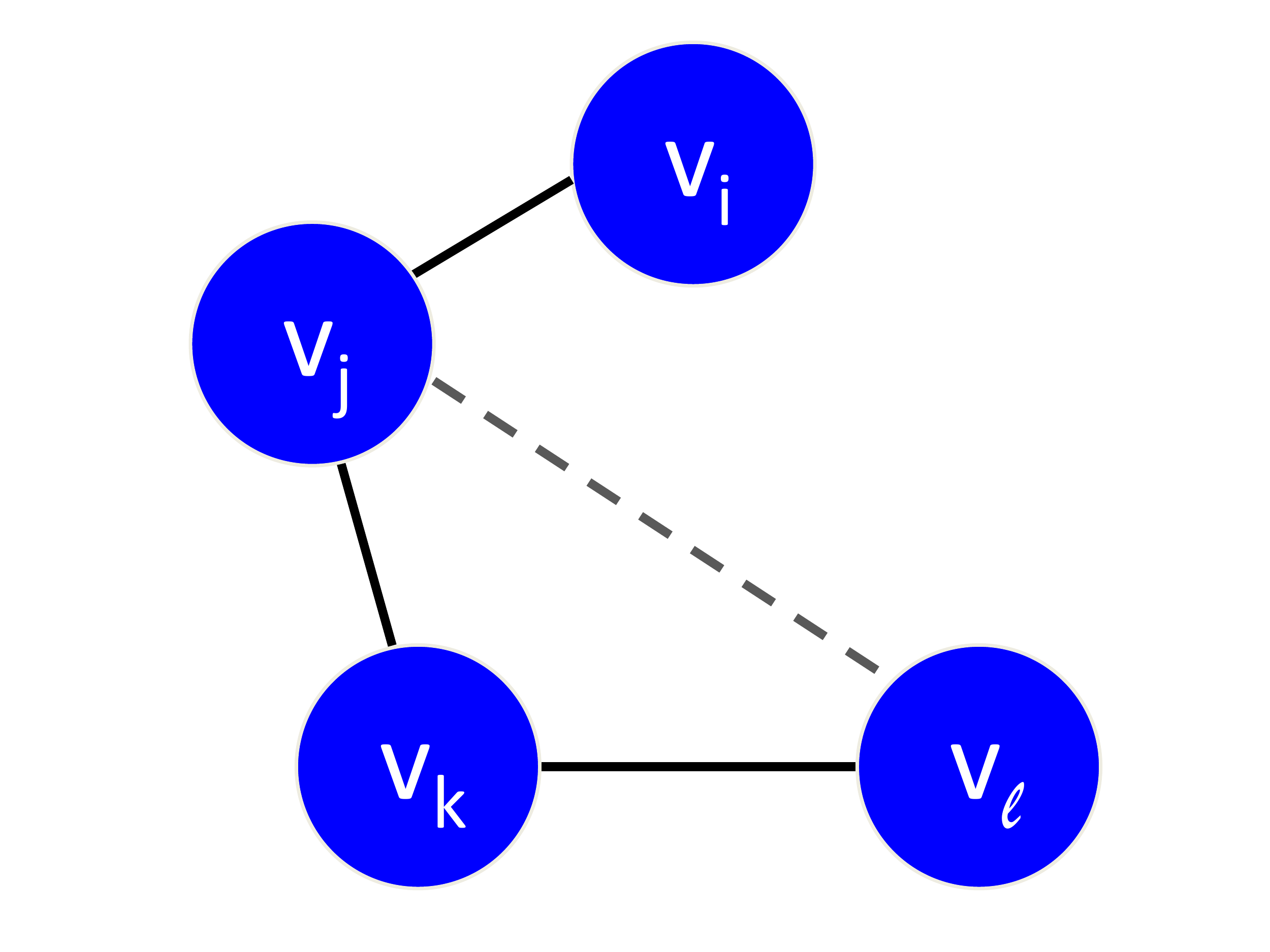}}
\subfigure{\includegraphics[width=.25\textwidth]{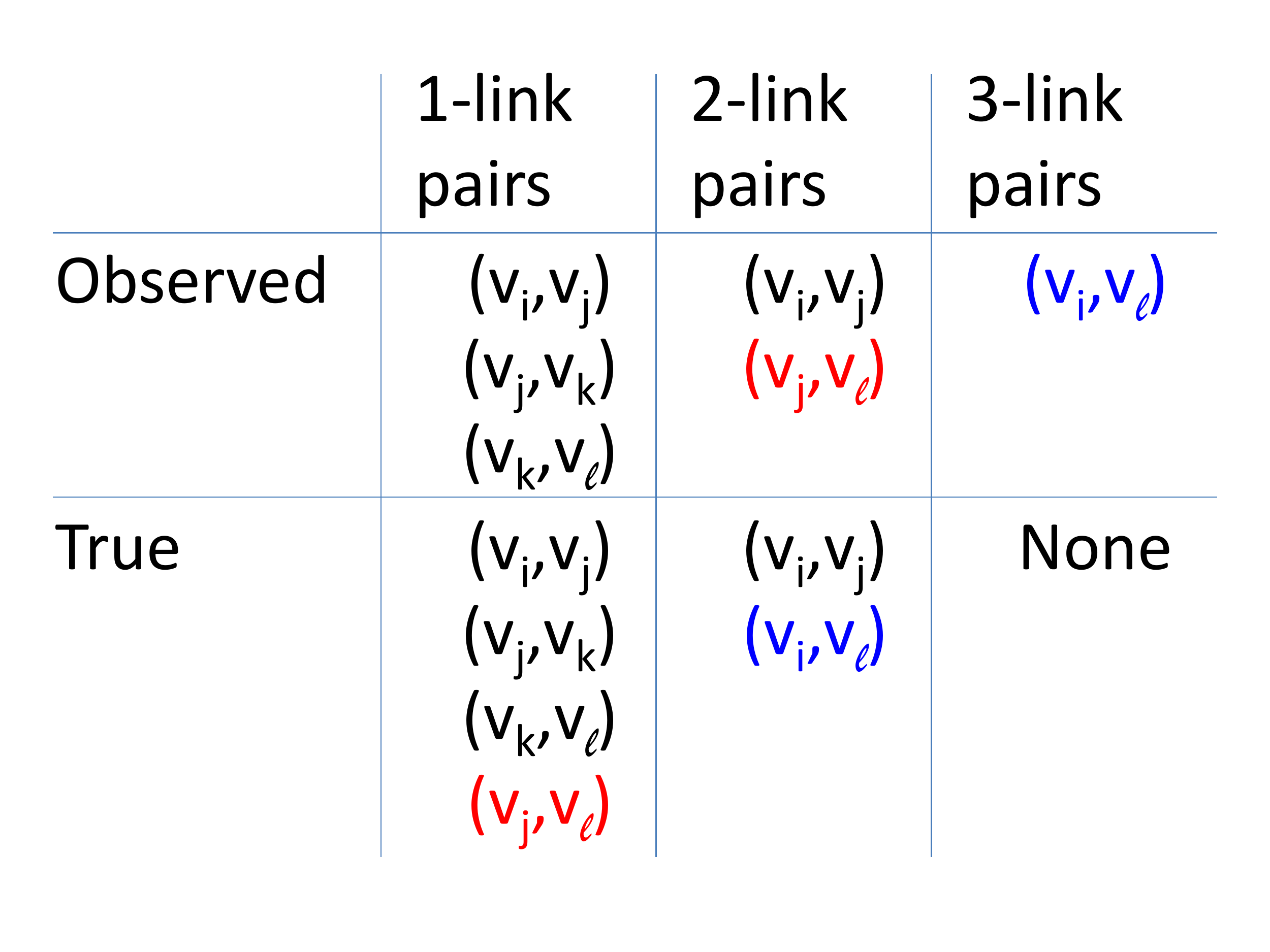}}
\caption{
The effect of missing links in the reciprocal reply network is depicted where observed links are shown as a solid line and an unobserved link is shown as a dashed line.
The effect of unobserved links is twofold:
(1) some connections between nodes are missed
(e.g., $v_j$ and $v_{\ell}$ are not connected in the observed reciprocal reply network);
and (2) some path lengths between nodes are artificially inflated
(e.g., the distance from $v_i$ to $v_{\ell}$ is 3 in the observed reciprocal-reply network,
however in reality the path length is 2).
}

\label{fig:missing_links_diagram}
     \end{figure}

We acknowledge two sources of missing data. First, the Twitter API did not allow us access to all tweets posted during the 6 month period under consideration. Thus, there are replies that we have not observed. As a result, some users may remain unconnected or connected by a path of longer length due to missing intermediary links in our reciprocal-reply network (Fig. \ref{fig:missing_links_diagram}). Secondly, we acknowledge that users may be interacting with each other and not using the built-in reply function. We discuss this further in the next section.

\subsection{Reciprocal-reply network}
In keeping with terminology used in the field of complex networks, the terms \textit{nodes} and \textit{links} will be used henceforth to describe users and their connections. Define $\mathcal{G}=(V,E)$ to be a simple graph which contains, $N=|V|$ nodes and $M=|E|$ links. We construct the reciprocal-reply networks in which users are represented by nodes, $v_i \in V$, and links connecting two nodes, $e_{ij} \in E$, indicate that $v_i$ and $v_j$ have made replies to each other during the period of time under analysis (Fig. \ref{fig:follower_bid_network}). For each network, we remove self-loops (i.e., users who responded to themselves).  
We characterize the reciprocal-reply network for each week by the calculation of network statistics such as $N$ (the number of nodes), $\left\langle k \right\rangle$ (average degree), $k_{\max}$ (maximum degree), the number of connected components and $S$ (proportion of nodes in the giant component). We calculate clustering, $C_G$, according to Newman's global clustering coefficient \cite{newman2001clustering}:  
\begin{equation*}
C_G=\frac{3 \times (\text{number of triangles on a graph})}{\text{number of connected triples of nodes}}. 
\end{equation*}

Assortativity refers to the extent to which similar nodes are connected in a network. Often, degree assortativity is quantified by computing the Pearson correlation coefficient of the degrees at each end of links in the network \cite{newman2002assortative}. Since we are interested in quantifying the extent to which the highest degree nodes are connected to other high degree nodes, as defined by the rank of their degrees, we instead measure degree assortativity by the Spearman correlation coefficient.\footnote{We present both the Spearman and Pearson correlation coefficient in the Appendix, Figure A2. Pearson's correlation coefficient is more sensitive to extreme values and thus obscures the trend in the data, namely that the network is assortative with respect to the rank (i.e., ordering) of nodes' degrees.} Thus for each link that connects nodes $v_i$ and $v_j$, we examine the ranks of $k_{v_i}$ and $k_{v_j}$. The Spearman correlation coefficient, which is the Pearson correlation coefficient applied to the ranks of the degrees at each end of links in the network, is a non-parametric test that does not rely on normally distributed data and is much less sensitive to outliers.\footnote{Our degree distribution is not Gaussian, as can be seen from Figure \ref{fig:prk_week}.} 

In addition, we also investigate user pairs which are connected by a minimal path length of two (or three) in the reciprocal reply networks. We define $d(v_i,v_j)$ to be the path length (i.e., number of links) between nodes $v_i$ and $v_j$ such that no shorter path exists. As a consequence of missing messages, we recognize that some users will appear to remain unconnected or connected by a path of longer length. Figure \ref{fig:missing_links_diagram} depicts the effect of missing links on inferred path lengths between nodes in the network. Nodes $v_j$ and $v_{\ell}$ are adjacent in the network, however, due to the missing link represented by the dashed line, these nodes are inferred to be two links apart.
\begin{figure}[htp!]
\centering     
            \includegraphics[width=.45\textwidth]{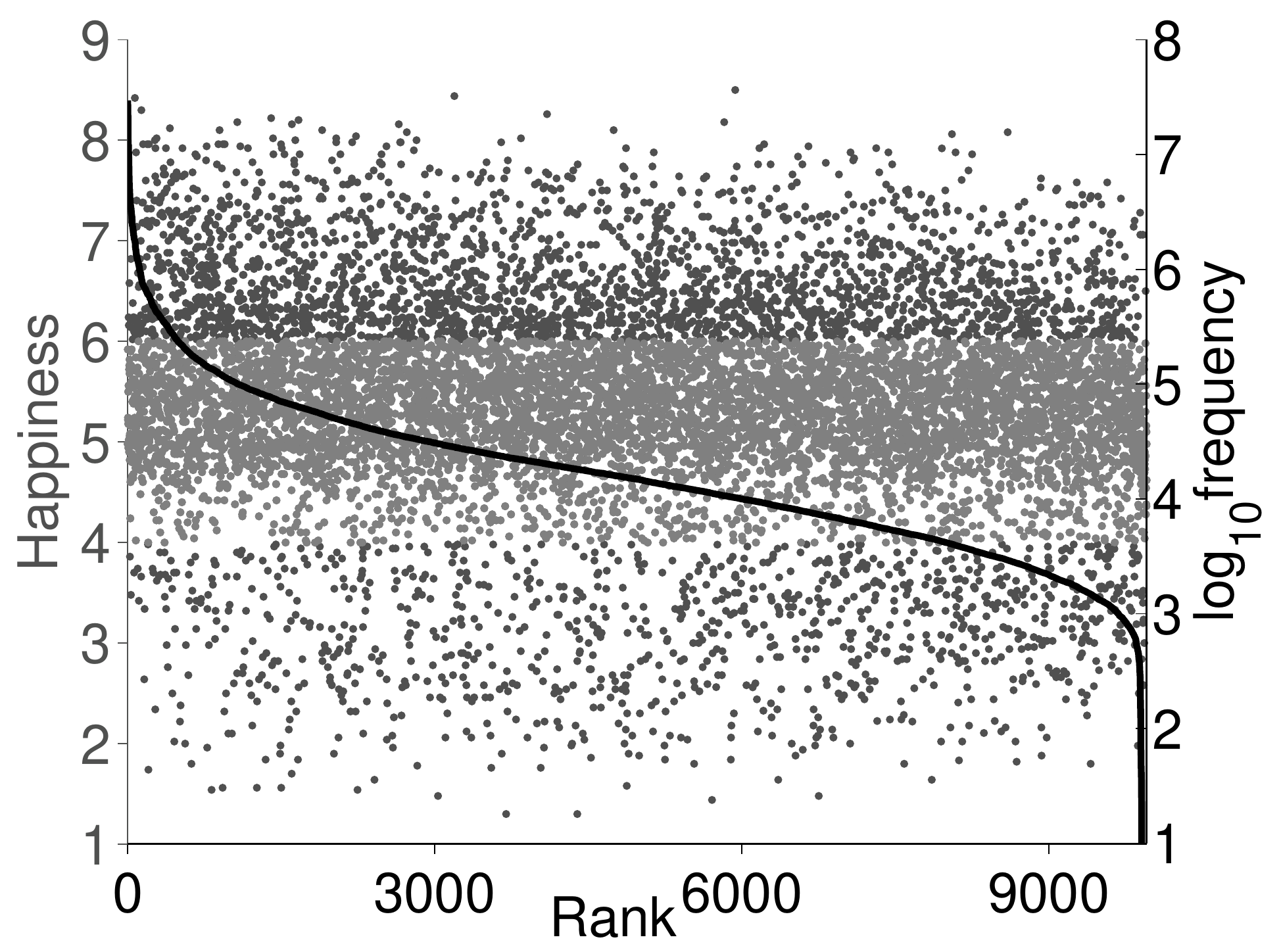}
\caption{The happiness scores of words are plotted as a function of their rank (dots), with the stop words (words within $\pm \Delta h=1$ of $h_{\text{avg}}=5)$ depicted in light grey \cite{Kloumann2011}. These words were excluded from the happiness score computation. The frequency of words and their rank (1=most frequent, 9956=least frequent) are plotted (solid curve). Not all 10,222 labMT words were observed during the time period from September 2008-February 2009.}
      \label{fig:rankvsfreq}
     \end{figure} 
\begin{figure*}[htp!]
\centering     
                      \includegraphics[width=\textwidth]{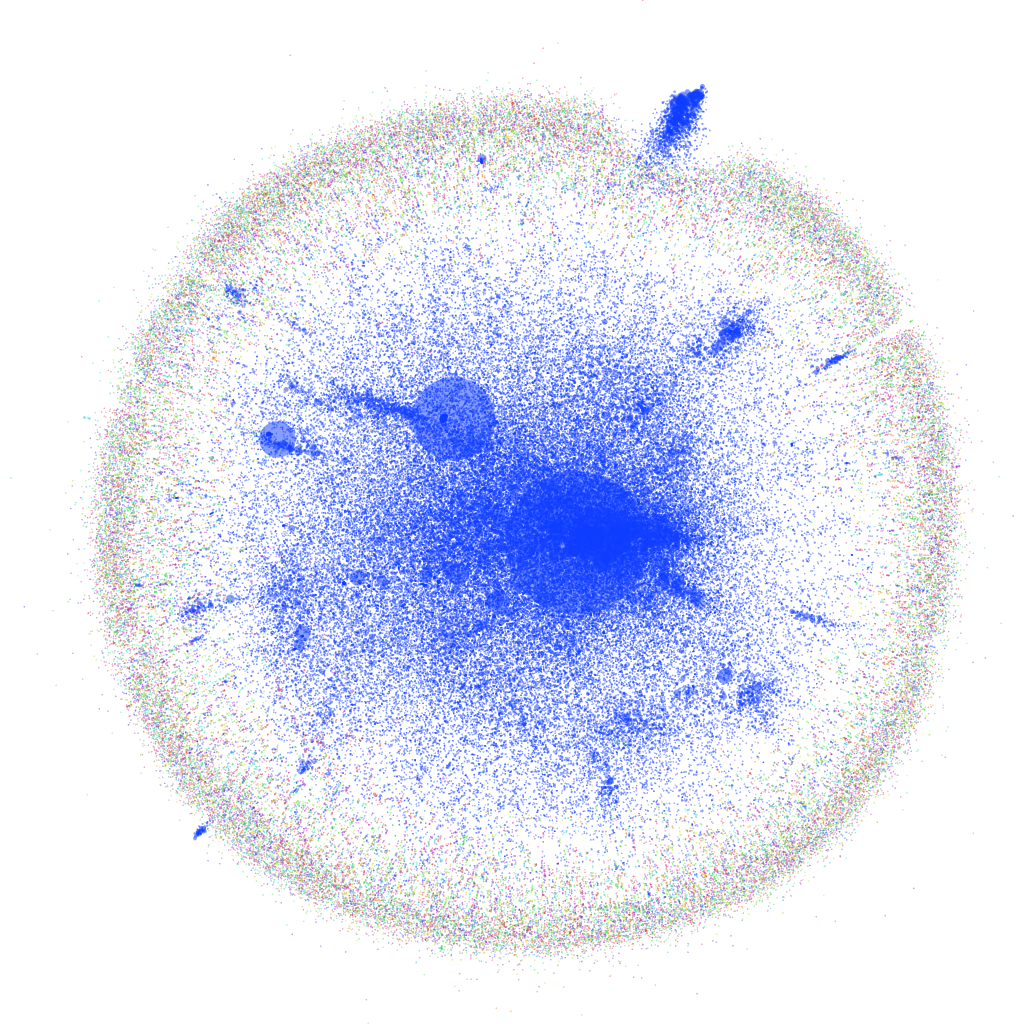}
\caption{A visualization of the 162,445 nodes in the reciprocal reply network for the week beginning December 9, 2008 (Week 14) is depicted. Node colors represent connected components, a total of 15342, with the giant component (shown in blue) comprising 76 \% of all nodes. The size of each node is proportional to its degree. The visualization was made using Gephi \cite{gephi}.}
      \label{fig:Giant_c}
     \end{figure*}      

     \begin{figure*}[htp!]{
\centering     
\subfigure{
\includegraphics[width=.32\textwidth]{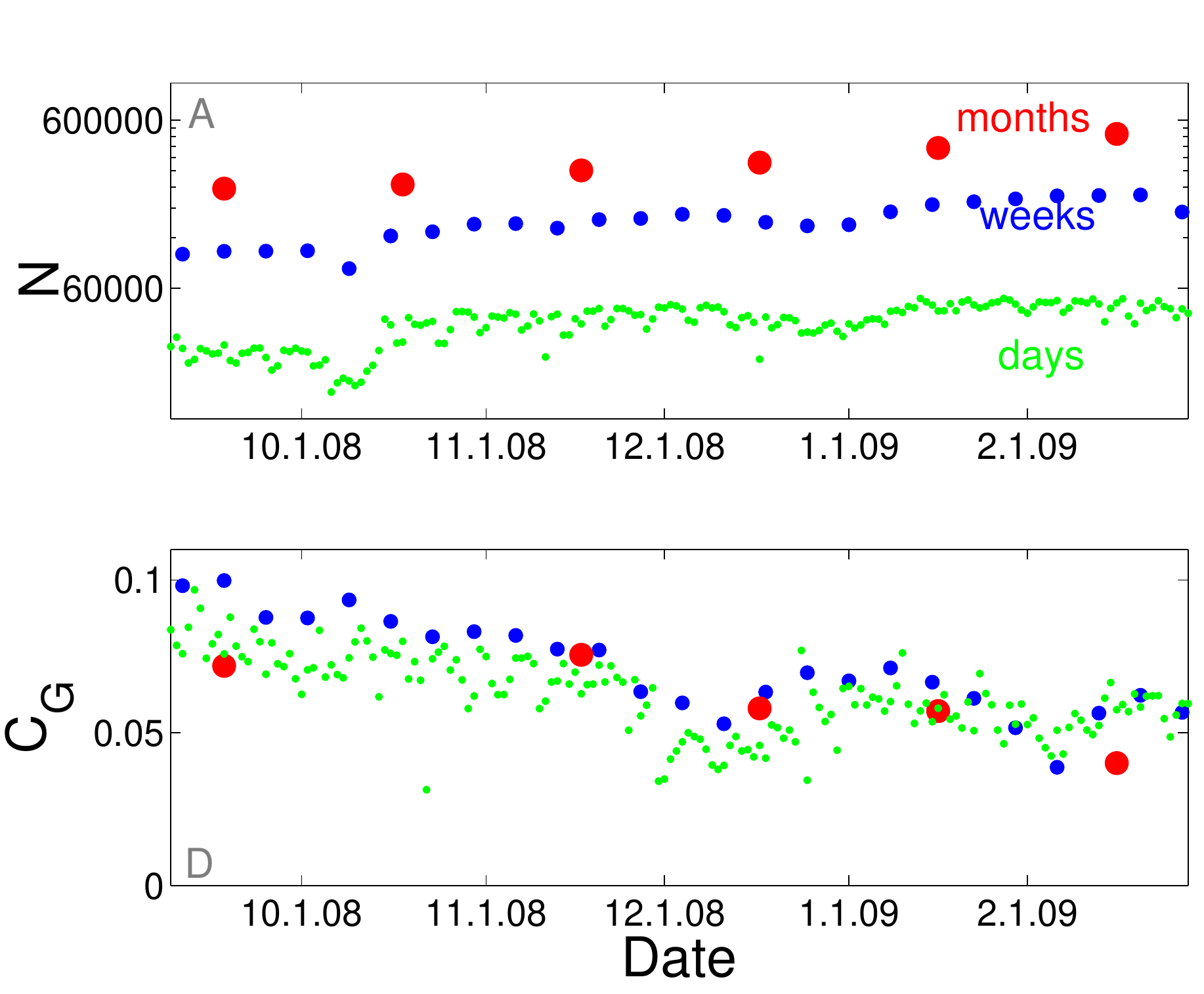} }
\subfigure{
 \includegraphics[width=.32\textwidth]{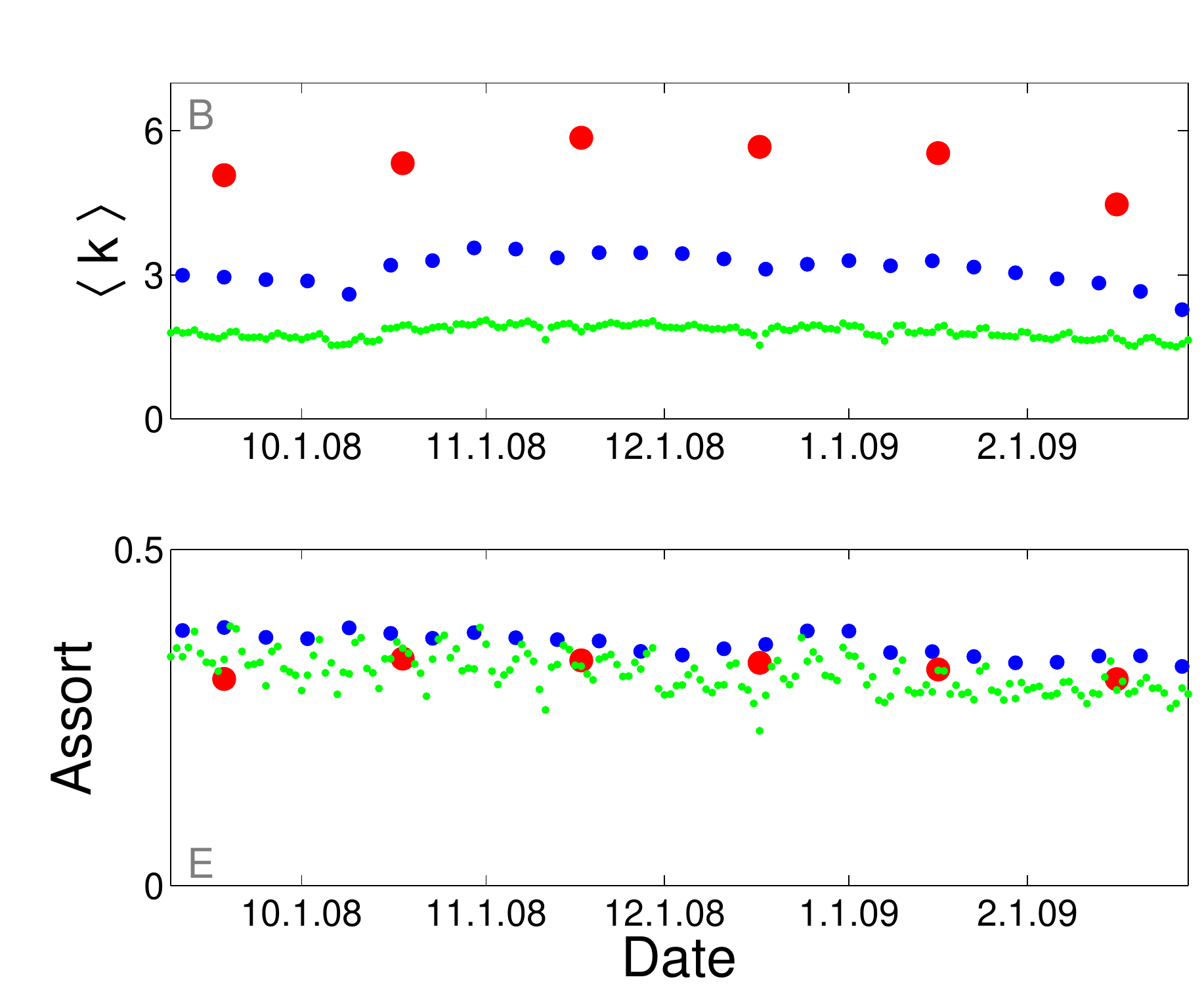} }
\subfigure{
\includegraphics[width=.32\textwidth]{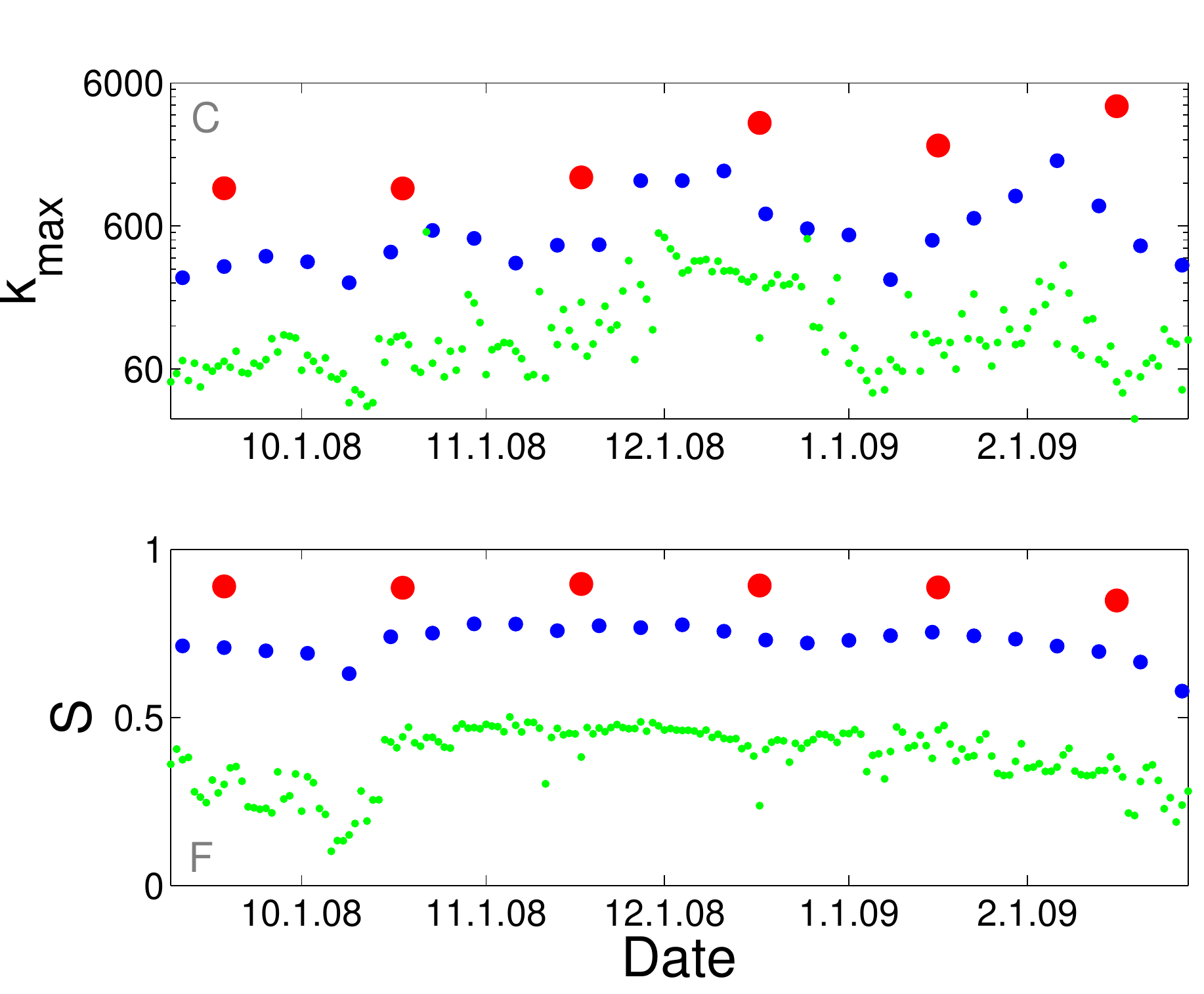} }
\caption{Network statistics for the reciprocal-reply network are constructed at the scale of days (green), weeks (blue), and months (red). (A.) The number of users ($N$) engaged in reciprocal exchanges when viewed at the level of days, weeks, or months increases over the study period. (B.) The average degree ($\left\langle k \right\rangle$) remains fairly constant throughout the study period, with higher values detected for larger interaction time periods. (C.) The maximum degree ($k_{\max}$) shows variability throughout the study period. (D.) Clustering decreases quite likely resulting from the inability of the networks' closed triangles to keep up with the growing number of nodes. (E.) Degree assortativity remains fairly constant throughout the study period, and shows little sensitivity to the time period over which the networks represent interactions. (E.) The proportion of nodes in the giant component ($S$) remains fairly constant for week and month networks, however, shows some variability during the first month of the study for day networks.}
\label{fig:net_stats}}
     \end{figure*} 

\subsection{Measuring happiness}
To quantify happiness for Twitter users, we apply the real-time hedonometer methodology for measuring sentiment in large-scale text developed in Dodds et al.~\cite{Dodds2011}. In this study, the 5000 most frequently used words from Twitter, Google Books (English), music lyrics (1960 to 2007) and the New York Times (1987 to 2007) were compiled and merged into one list of 10,222 unique words.\footnote{We provide a brief summary of this methodology here and refer the interested reader to the original paper for a full discussion. The supplementary information contains the full word list, along with happiness averages and standard
deviations for these words  \cite{Dodds2011}.} This word list was chosen solely on the basis of frequency of usage and is independent of any other presupposed significance of individual words. Human subjects scored these 10,222 words on an integer scale from 1 to 9 (1 representing sad and 9 representing happy) using Mechanical Turk. We compute the average happiness score ($h_{\text{avg}}$) to be the average score from 50 independent evaluations. Examples of such words and their happiness scores are: $h_{\text{avg}}$(love)=8.42, $h_{\text{avg}}$(special)=7.20, $h_{\text{avg}}$(house)=6.34, $h_{\text{avg}}$(work)=5.24, $h_{\text{avg}}$(sigh)=4.16, $h_{\text{avg}}$(never)=3.34, $h_{\text{avg}}$(sad)=2.38, $h_{\text{avg}}$(die)=1.74. Words that lie within $\pm \Delta h_{\text{avg}}=1$ of $h_{\text{avg}}$=5 were defined as ``stop words'' and excluded to sharpen the hedonometer's resolution.\footnote{For notational convenience, we henceforth use $\Delta h$ in lieu of $\Delta h_{\text{avg}}$.}
The result is a list of 3,686 words, hereafter referred to as the Language Assessment by Mechanical Turk (labMT) word list \cite{Dodds2011}. See Tables A1 and A2 for additional example word happiness scores. 

\begin{table}[ht!]
\label{table_anew_stats}
\begin{center}
\begin{tabular}{ccccc}
\hline
$w_i$  & $h_{\text{avg}}(w_i)$ & labMT? & $f_i$ & $p_i$   \\ 
\hline 
\hline
Vacation & 7.92 & yes & 1 & $\frac{1}{2}$     \\ 
starts & 5.96 & yes  & n/a & n/a   \\ 
today  &  6.22  & yes  & 1 & $\frac{1}{2}$     \\ 
yeahhhhh & n/a & no       & n/a & n/a\\
\hline
\end{tabular}
\end{center}
\caption{Happiness scores are computed as a weighted average of words' $h_{\text{avg}}$ scores. Since ``starts'' is a stop word, it is not included in the calculation of $h_{\text{avg}}(T)=7.07$. This example serves is included as a means to illustrate the methodology; in practice, the average is calculated over a much larger word set.}
\end{table}
     
Figure \ref{fig:rankvsfreq} presents word happiness as a function of usage rank for the roughly 10,000 words in the labMT data set. This figure reveals a frequency independent bias towards the usage of positive words (see [37] for further discussion of this positivity bias). Proceeding with the labMT word list, a pattern-matching script evaluated each tweet for the frequency of words. We compute the happiness of each user by applying the hedonometer to the collection of words from all tweets authored by the user during the given time period. Note that each users' collection of words likely reflects messages that were not replies. The happiness of this collection of words is taken to be the frequency weighted average of happiness scores for each labMT word as $h_{\text{avg}}(T) =\frac{\sum^N_{i=1} h_{\text{avg}}(w_i)f_i }{\sum^N_{i=1} f_i}=\sum^N_{i=1} h_{\text{avg}}(w_i) p_i,$ where $h_{\text{avg}}(w_i)$ is the average happiness of the $i$th word appearing with frequency $f_i$ and where $p_i$ is the normalized frequency ($p_i=\frac{f_i}{\sum^N_{j=1} f_j}$). As a simple example example, we consider the phrase: \textit{Vacation starts today, yeahhhhh!} in Table 1. In practice, though, the hedonometer is applied to a much larger word set and is not applied to single sentences.

Having found happiness scores for each node (user), we then form happiness-happiness pairs  $(h_{v_i},h_{v_j})$, where $h_{v_i}$ and $h_{v_j}$ denote the happiness of nodes $v_i$ and $v_j$ connected by an edge. The Spearman correlation coefficient of these happiness-happiness pairs measures how similar individuals' average happiness is to that of their nearest neighbors'. Lastly, we investigate the strength of the correlation between users' average happiness scores and those of other users in the two and three link neighborhoods.  

\vspace{-3mm}
\section{Results}
\vspace{-3mm}
\subsection{Reciprocal-reply network statistics}
Visualizations of day and week networks were created using the software package Gephi \cite{gephi}. Figures \ref{fig:Giant_c} and \ref{fig:day_modularity} show a sample week and day network, respectively. All layouts were produced using the Force Atlas 2 algorithm, which is a spring based algorithm that plots nodes together if they are highly connected (see \cite{forceatlas2} for more details). The sizes of the nodes are proportional to the degrees.

Network statistics, such as the number of nodes (\textit{N}), the average degree $\left\langle k \right\rangle$, the maximum degree ($k_{\max}$), global clustering $C_G$, degree assortativity ($Assort$), and the proportion of nodes in the giant component ($S$) are summarized in Figure \ref{fig:net_stats}.
Several trends are apparent.
\begin{figure}[htp!]{
\centering     
\includegraphics[width=.45\textwidth, height=6.8cm]{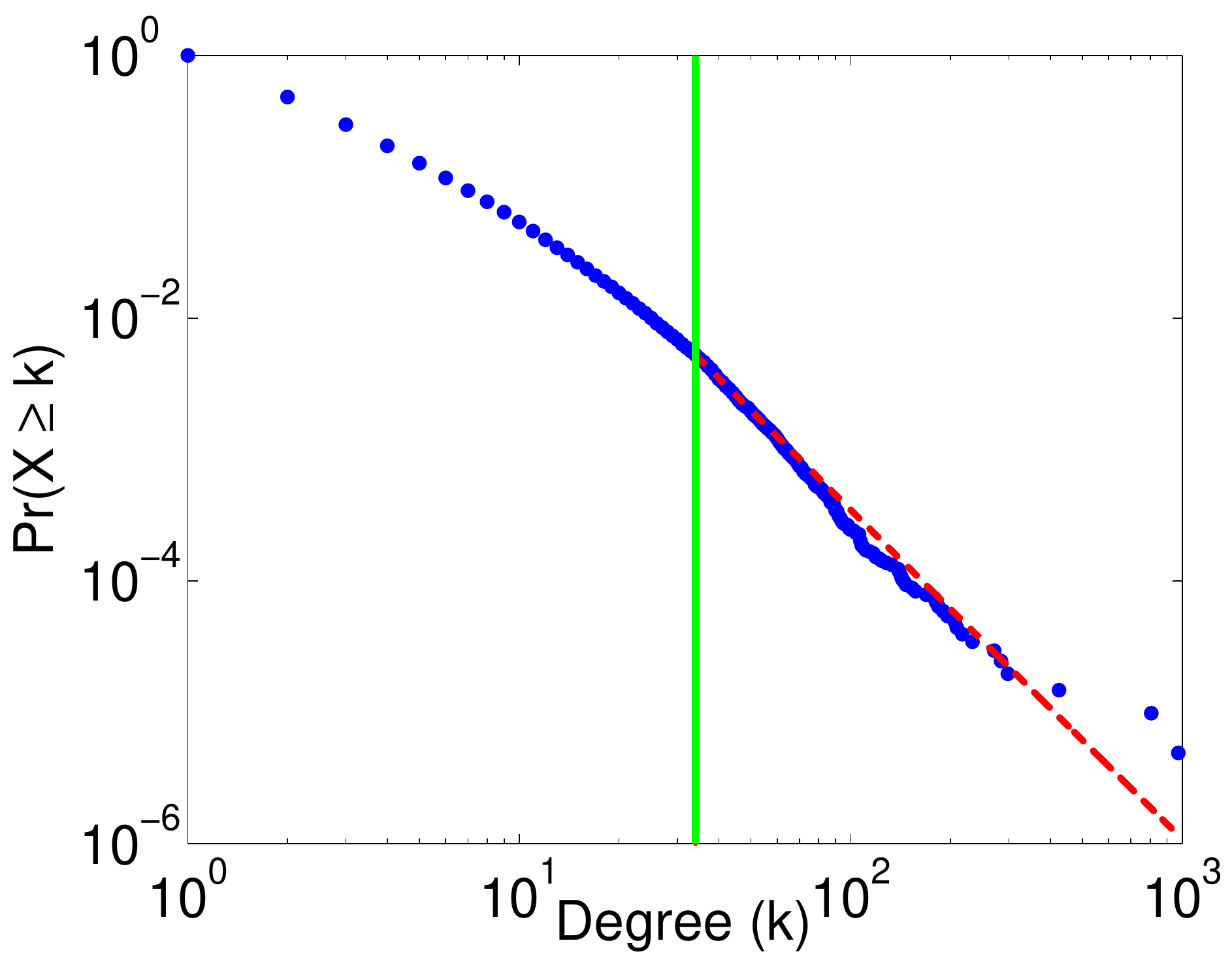}
\caption{Log-log plot of the complementary cumulative distribution function (CCDF) of the degree distribution for a sample week (week of January 27, 2009) network is shown (blue), along with the best fitting power law model ($\alpha=3.50$ and $k_{\min}=34$) using the procedure of Clauset, Shalizi, and Newman \cite{clauset_powerlaw}. We test whether the empirical distribution is distinguishable from a power law using the Kolmogorov-Smirnov test and find no evidence against the null hypothesis ($D=2.28 \times 10^{-2} , p=0.095, n=203852$).}
\label{fig:prk_week}}
     \end{figure}

Throughout the course of the study, the number of users in the observed reciprocal-reply network shows an increase, whereas the average degree, degree assortativity, and proportion of nodes in the giant component remain fairly constant. The fluctuations in maximum degree are the result of celebrities or companies having bursts of high volume reply exchanges with their fans during a particular week, for example Bob Bryar, Drummer for the band \textit{My Chemical Romance} ($k_{\max}=1244$, Week 12), \textit{Namecheap} domain registration company ($k_{\max}=1245$, Week 13), Twitter's own \textit{Shorty Awards} ($k_{\max}=1456$, Week 14), and Stephen Fry, actor and mega-blogger ($k_{\max}=1718$, Week 22). This observation highlights the importance of examining network data on the appropriate time scale, otherwise information about these kinds of dynamics would be be lost. The clustering coefficient shows a slight decrease over the course of this period. This is most likely due to an increasing number of nodes, which results in a smaller proportion of closed triangles in the network. 
 
The degree distribution, $P_k$, for a sample week (week beginning January 27, 2009) is presented in Figure \ref{fig:prk_week}. Using the approach outlined by Clauset, Shalizi, and Newman \cite{clauset_powerlaw}, we find a lower bound for the scaling region to be $k_{\min} \approx 34$ and a very steep scaling exponent of $\alpha=3.5$. This suggests a contrained variance and mean. We test whether the empirical distribution is distinguishable from a power law using the Kolmogorov-Smirnov test and find no evidence against the null hypothesis for the week ($D=2.28 \times 10^{-2}, p=0.095, n=203852$). We find the same exponent and statistically stronger evidence of a power law for a sample month (see the Appendix, Fig. \ref{fig:prk_month}). This suggests that these distributions' tails may be fit by a power law.

\subsection{Measuring happiness}
\begin{figure}[htp!]
\centering     
{\includegraphics[width=.45\textwidth, height=6.8cm]{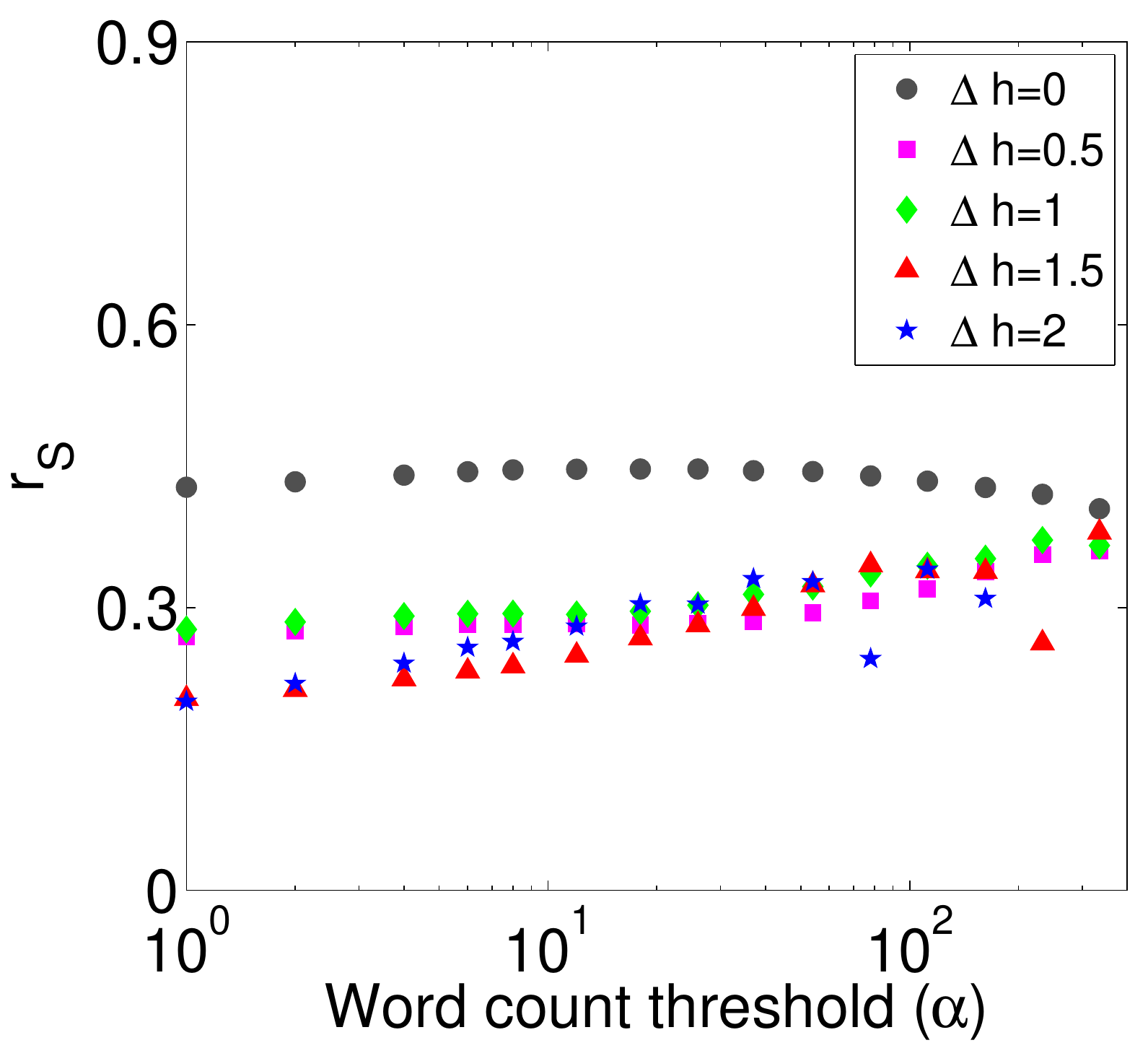}
      \caption{Nearest neighbor happiness assortativity as a function of the number of labMT words required per user is displayed for a sample week reciprocal-reply network. Notice that when $\Delta h=0$, there is less variation due to the numerous words centered around the mean happiness score regardless of the threshold, $\alpha$. While this stability is desirable, tuning $\Delta h$ allows us to sharpen the resolution of the hedonometer. This tuning, however, must be balanced with the appropriate choice of $\alpha$. 
      }
      \label{fig:progression_week}}
      \end{figure}

The application of the hedonometer gives reasonable results when applied to a large body of text, but can be misleading when applied to smaller units of language \cite{Dodds2011}. To provide a sense of how sensitive this measure is to the number of labMT words posted by users, we sampled happiness-happiness pairs, $(h_{v_i},h_{v_j})$ whose respective users, $v_i$ and $v_j$, had posted at least $\alpha$ total labMT words during a sample week (week beginning January 27, 2009). For these users, we compute happiness assortativity and show the variation with $\alpha$ in Figure \ref{fig:progression_week}. For $\Delta h=0$, there is less variation due to the numerous words centered around the mean happiness score regardless of the threshold, $\alpha$. %The same analysis for a sample month (January 2009) is presented in the Appendix (Fig. \ref{fig:progression_month}).
Tuning both parameters too high results in few sampled words and corrupts the interpretation of the results.

\begin{figure}[htp!]
\centering     
            \includegraphics[width=.45\textwidth]{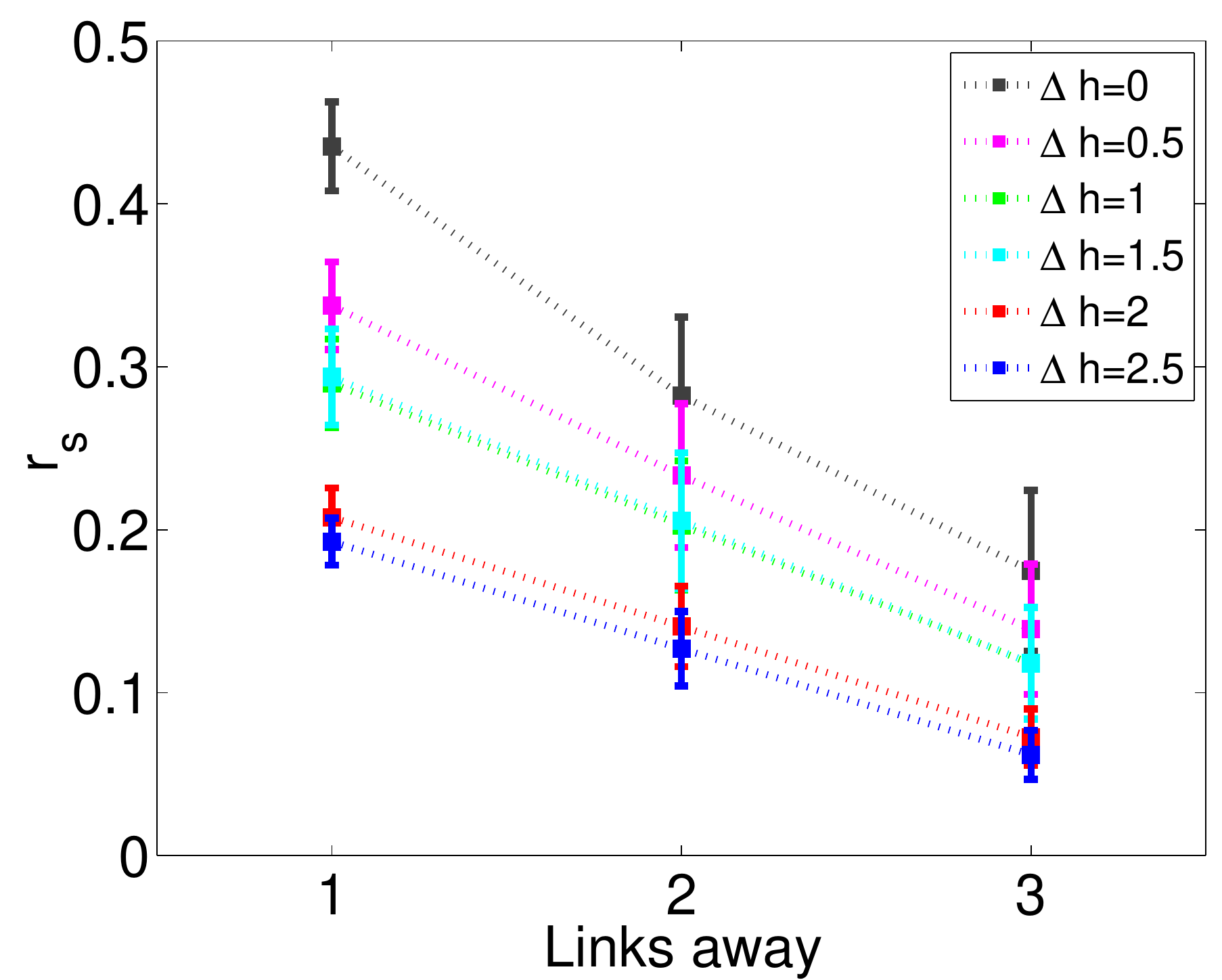}
                        
      \caption{Average assortativity of happiness for week networks measured by Spearman's correlation coefficients as $\Delta h$ is dialed from 0 to 2.5, with $\alpha=50$. As $\Delta h$ increases, the average correlation decreases. For large $\Delta h$ the resulting words under analysis have more disparate happiness scores and thus the correlations between users' happiness scores are smaller. Similarly, choosing $\Delta h$ to be too small (e.g., $\Delta h=0$) could result in an over estimate of happiness-happiness correlations because of the uni-modal distribution of $h_{\text{avg}}$ for the labMT words. Thus a moderate value for $\Delta h$ is chosen ($\Delta h$ is set to 1 for this study).}
      \label{fig:dial_deltah_spearman_links_away50}
      \end{figure}      

\begin{figure*}[htp!]
\centering     
\subfigure[$\Delta h=1$, $\alpha=1$]{
            \includegraphics[width=.47\textwidth, height=55mm]{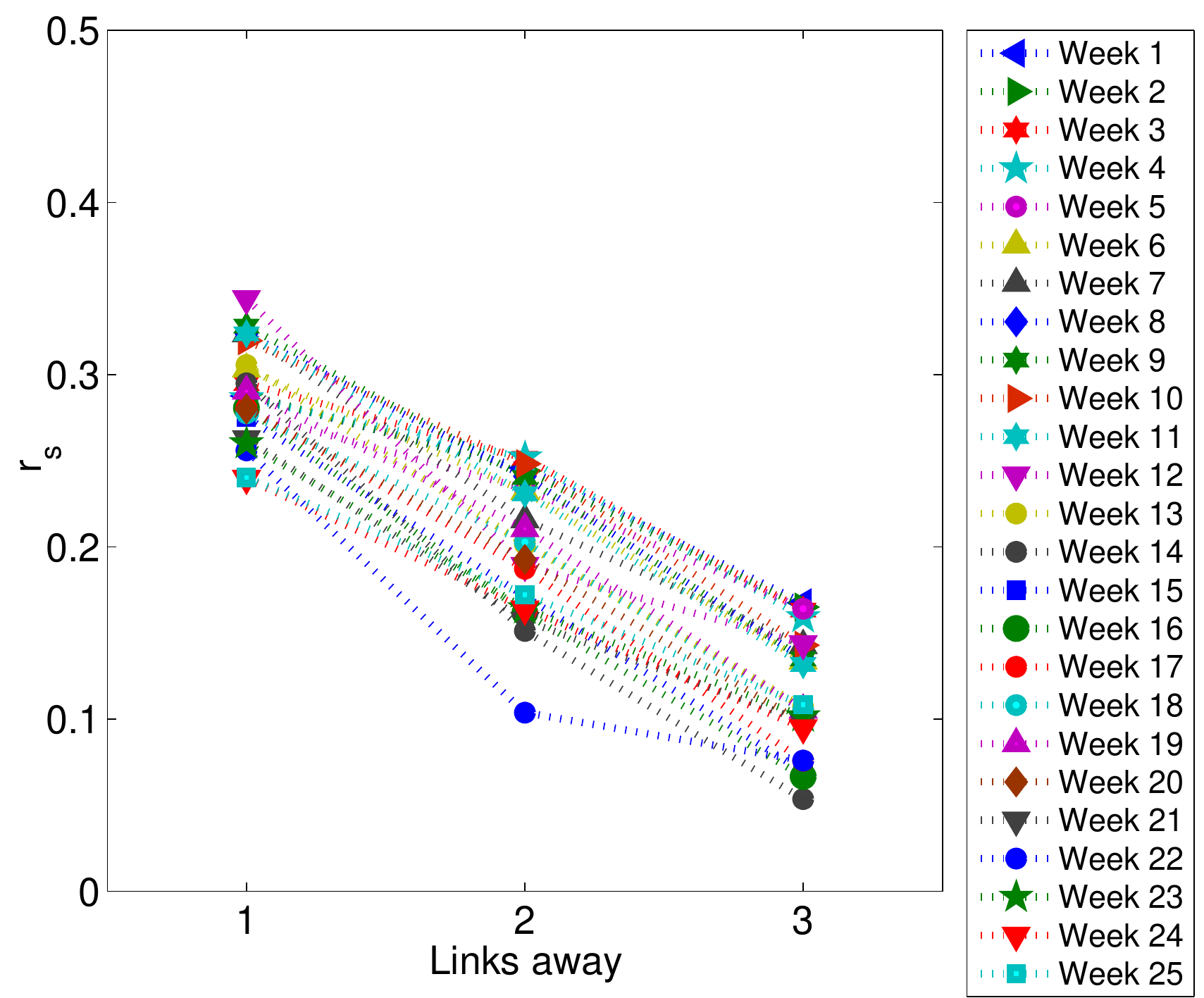}}
\subfigure[$\Delta h=1$, $\alpha=50$]{
            \includegraphics[width=.43\textwidth, height=55mm]{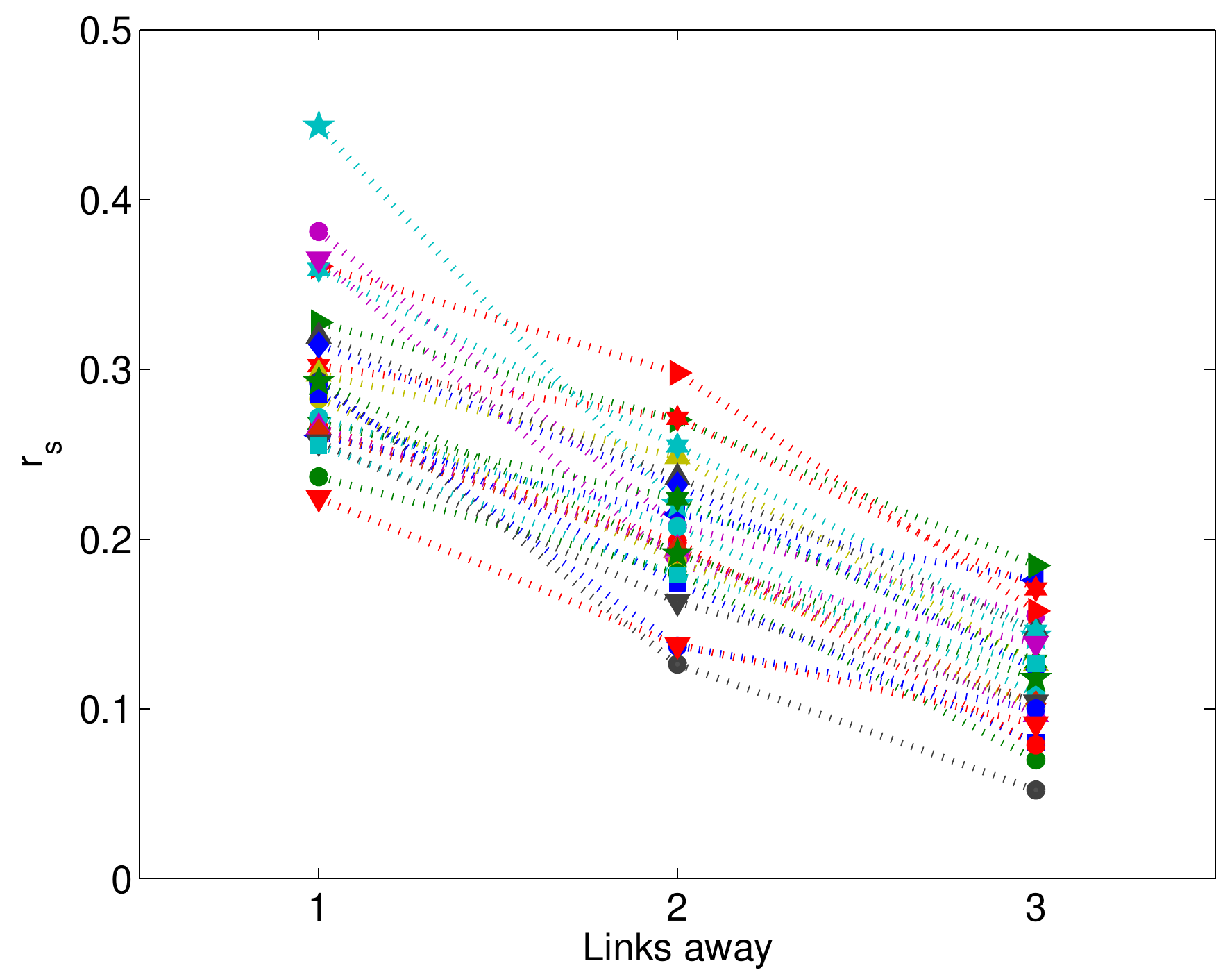}}    
      \caption{Happiness assortativity as measured by 
  Spearman's correlation coefficients is shown for week networks, with $\Delta h=1$ and (a) the threshold of labMT words written by users set to $\alpha=1$ and (b) $\alpha=50$. The dashed lines indicate weakening happiness-happiness correlations as the path length increases from one, two, and three links away, for each week in the data set. }
      \label{fig:linksaway}
      \end{figure*}
      
Figures \ref{fig:dial_deltah_spearman_links_away50} and \ref{fig:linksaway} reveal a weakening happiness-happiness correlation for users in the week networks as the path length between nodes increases. All correlations, for each week, were significant ($p<10^{-10}$). This suggests that the network is assortative with respect to happiness and that user happiness is more similar to their nearest neighbors than those who are 2 or 3 links away. 

\begin{figure}[htp!]
\centering           
                        \subfigure{\includegraphics[width=.45\textwidth]{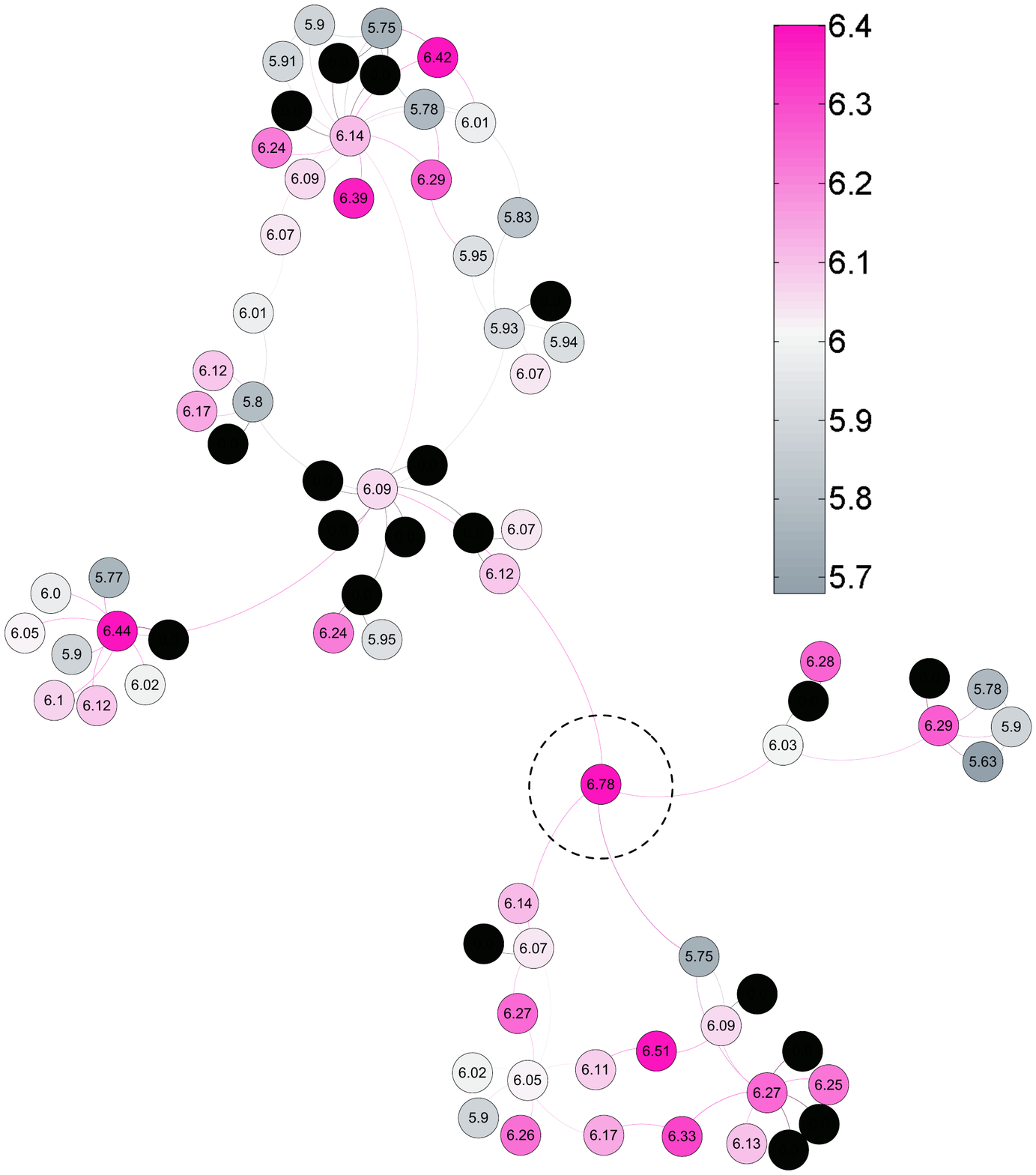}}
\caption{A visualization of a user and its neighbors 3-links away for a week beginning September 9, 2008 (Week 1). Colors represent happiness scores for users posting more than $\alpha=50$ labMT words. Nodes depicted with the color black are nodes for which the user's wordbag did not meet our thresholding criteria. }
      \label{fig:sample_3link}
     \end{figure}

In Figure \ref{fig:sample_3link} we provide a visualization of an ego-network for a single node, including neighbors up to three links away.  Nodes are colored by their $h_{\text{avg}}$ score, illustrating the assortativity of happiness.  Figure \ref{fig:pighate} visualizes the happiness assortativity for an entire week network.

\begin{figure}[htp!]{    
\includegraphics[width=.48\textwidth]{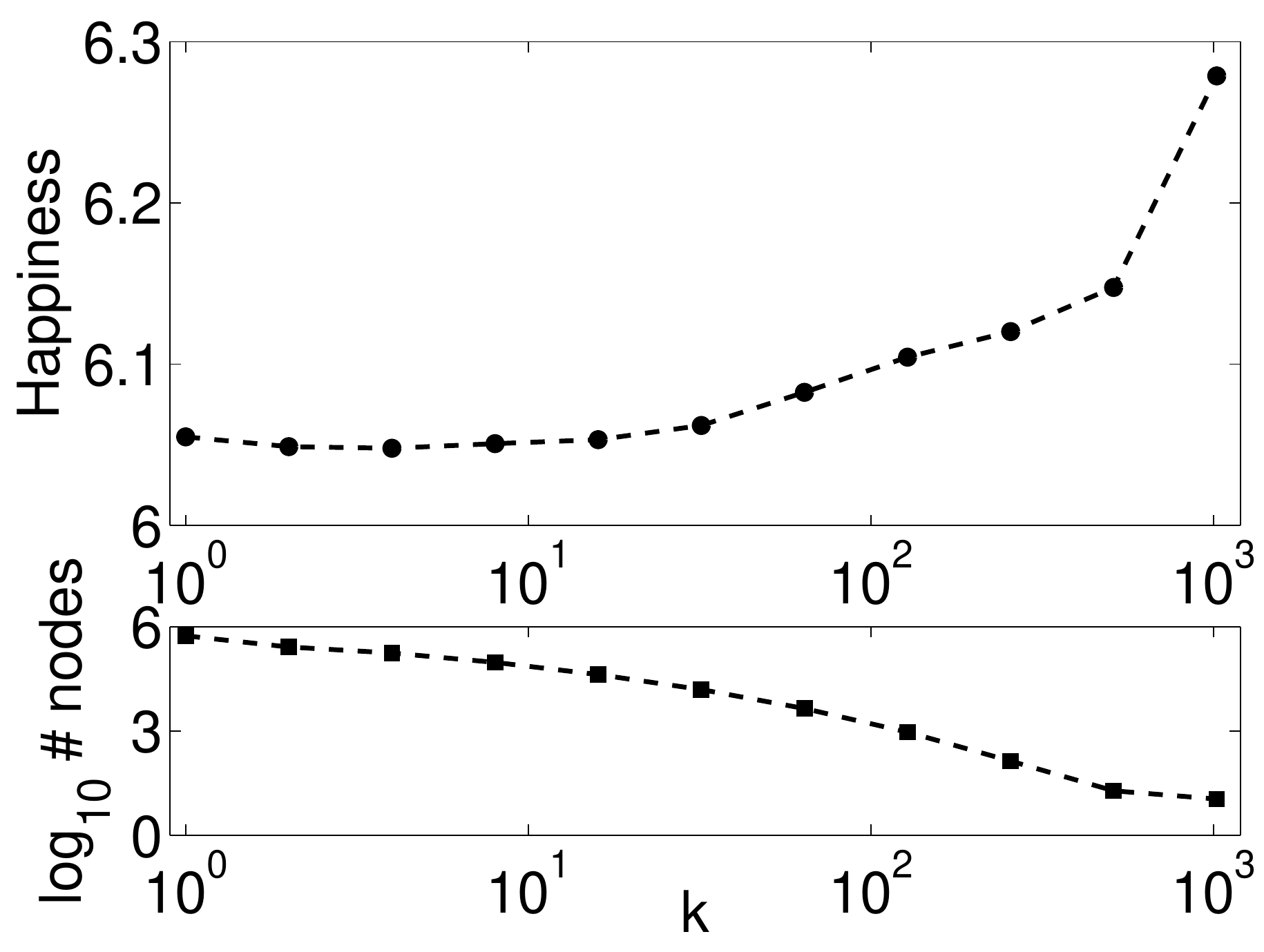}
\caption{
Top Panel: The average happiness score  as a function of user degree $k$ for week networks is increasing, as larger degree nodes use fewer negative words (see Figure \ref{fig:worshift}). Bottom Panel: The number of unique users is reported with respect to degree $k$; some users appear in more than one bin because they exhibit different degree $k$ for different weeks of the study.}
\label{fig:deg_vs_happiness_wordbag}}
     \end{figure}

In Figure \ref{fig:deg_vs_happiness_wordbag}, we show the average happiness score as a function of user degree $k$ for all week networks. The average happiness score increases gradually as a function of degree, with large degree nodes demonstrating a larger average happiness than small degree nodes. Large degree nodes use words such as ``you,'' ``thanks,'' and ``lol'' more frequently than small degree nodes, while the latter group uses words such as ``damn,'' ``hate,'' and ``tired'' more frequently. A word shift diagram, comparing nodes with $k<100$ vs. nodes with $k \geq 100$ is included in the Appendix (Fig. \ref{fig:worshift}). Figure \ref{fig:deg_vs_happiness_wordbag} also reveals that the number of large degree nodes is fairly small. Our results support recent work showing that most users of Twitter exhibit an upper limit on the number of active interactions in which they can be engaged \cite{Goncalves2011}. This may provide further evidence in support of Dunbar's hypothesis, which suggests that the number of meaningful interactions one can have is near 150 \cite{Dunbar1992}.    

      \subsection{Testing assortativity against a null model}
\begin{figure}[htp]
\centering 
      \includegraphics[width=.45\textwidth, height=55mm]{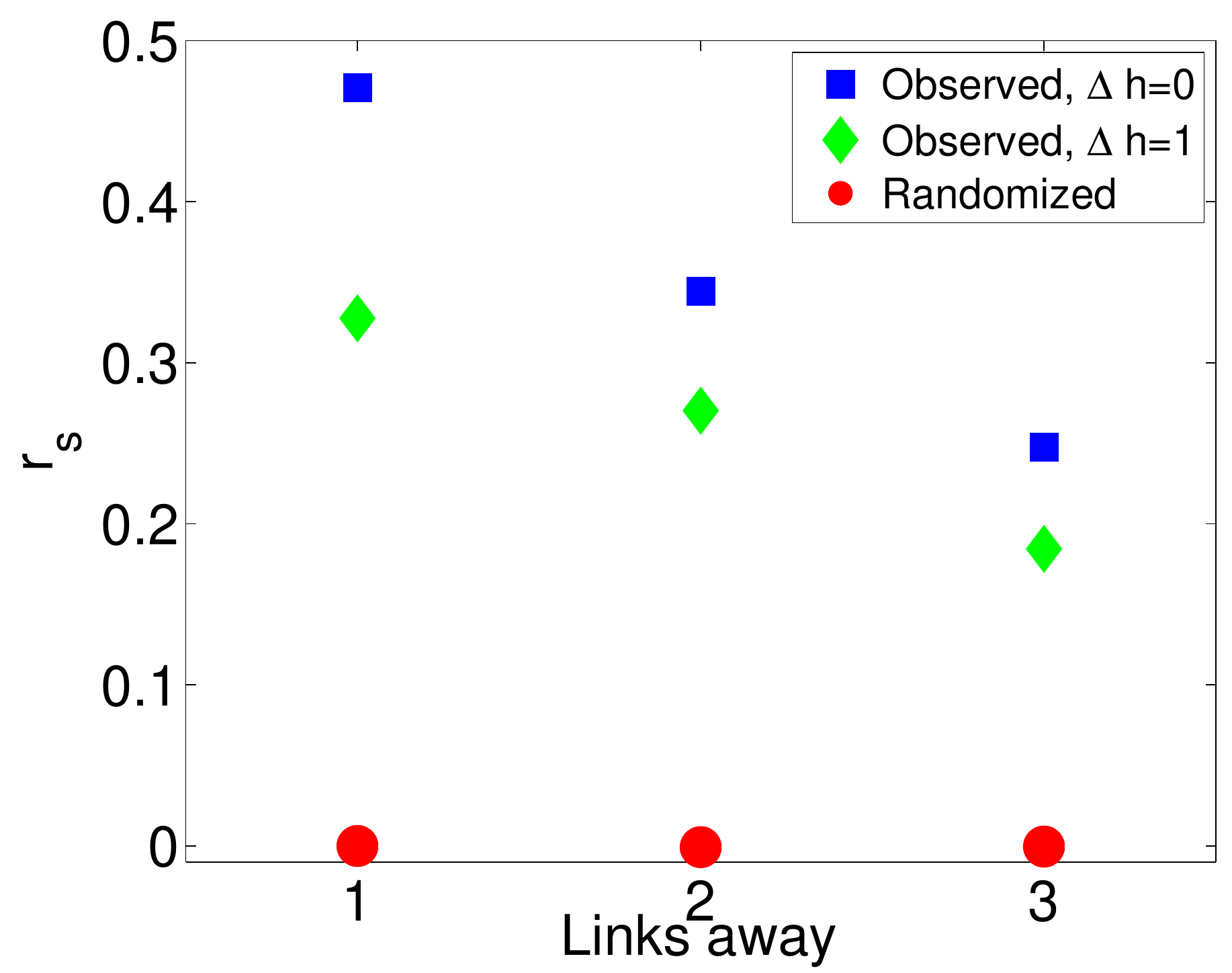}
\caption{One hundred random permutations were applied to the happiness scores associated with each node in a sample week network (week beginning October 8, 2008 is shown), with $\Delta h=0$ (blue square) and $\Delta h=0$ (green diamonds). The threshold for all cases is set to $\alpha=50$. The Spearman correlation coefficients, $r_s$ for the observed data are shown as blue squares. The average and standard deviation of the Spearman correlation coefficient calculated for the 100 randomized data (null model) are shown as red circles with error bars (the error bars are smaller than the symbol). The plot shows Spearman correlation coefficients for the null model to be nearly 0 and provides supporting evidence for our observed trend, namely the network is assortative with respect to happiness and the strength of assortativity decreases as path length increases.}
     \label{fig:rvalues}
     \end{figure}

To further examine these findings, we create a null model which maintains the network topology (i.e., adjacency matrices for one link, two link, and three link remain intact), but randomly permutes the happiness scores associated with each node. The Spearman correlation coefficient shows no statistically significant relationship for the null model applied to a sample week of the data set.  Figure \ref{fig:rvalues} shows the results of 100 random permutations applied to nodes' associated happiness scores. The Spearman correlation coefficients for the observed data are shown as blue squares ($\Delta h_{\text{avg}}=0$) and green diamonds ($\Delta h_{\text{avg}}=1$). The average and standard deviation of the Spearman correlation coefficient calculated for the 100 randomized happiness scores (null model) are shown as red circles with error bars (the error bars are smaller than the symbol). This data supports the hypothesis that happiness is less assortative as network distance increases. 

Lastly, we explore whether these correlations are due to similarity of word usage. For this analysis, we compute the similarity of word bags for users connected in the reciprocal reply networks. We compare the distribution of observed similarity scores to similarity scores obtained by randomly reassigning word bags to users. Figure \ref{fig:shuffled_hamming} shows that both distributions are of a similar form, with the randomized version exhibiting a slightly lower mean similarity score ($\overline{D_{i,j}}=.167)$ as compared to the mean of the observed similarity scores for users ($\overline{D_{i,j}}=.267)$. If users were tweeting similar words with a similar frequency, we would expect a much larger mean similarity score for the observed data. Thus, we do not find evidence suggesting that the happiness correlations are due to similarity of word bags.

\section{Discussion}

In this paper, we describe how a social sub-network of Twitter can be derived from reciprocal-replies. Countering claims that Twitter is not social a network \cite{Kwak2010}, we provide evidence of a very social Twitter. The large volume of replies (millions every week) and assortativity of user happiness indicates that Twitter is being used as a social service. Furthermore, conducted at the level of weeks, our analysis examines an in the moment social network, rather than the stale accumulation of social ties over a longer period of time. A network in which edges are created and never disintegrate results in dead links with no contemporary functional activity. This problem of unfriending has been noted \cite{noel2011} and can greatly impact conclusions drawn when observational data are used to infer contagion.

Our characterization of the reciprocal reply network reveals several trends over the 25 week period from September 2008 to February 2009. The number of nodes, $N$, in a given week network increased as time progressed, which is undoubtedly due to Twitter's enormous growth in popularity over the study period. Similarly, with an increasing number of nodes, we observe a smaller proportion of closed triangles (i.e., clustering shows a slight decrease). This may be due in part to sub-sampling effects or due to an increasing $N$, with which the number of closed triangles (i.e., friends of friends) cannot keep up. The proportion of nodes in the giant component remains fairly constant, as does degree assortativity as measured by Spearman's correlation coefficient. Had we used the Pearson correlation coefficient, degree assortativity would have been highly variable (Fig. A1) due to the extreme values of maximum degree ($k_{\max}$) during weeks 12-14 and 22. Using the Spearman rank correlation coefficient, which is less sensitive to extreme values, we find that the degree assortativity is fairly constant.

Our work is based on a sub-sample of tweets and is thus subject to the effects of missing data. The problem of missing data has been addressed by several researchers investigating the impact of missing nodes \cite{Stumpf2005, Sadikov2011, Leskovec2006sampling, Lee2006, frantz2009robustness}, missing links, or both \cite{kossinets2006effects}. More specifically, the work of Stumpf \cite{Stumpf2005} shows that sub-sampled scale-free networks are not necessarily themselves scale-free. Further work which addresses the problem of missing messages and identifies the consequences of missing data on inferred network topology is needed to more fully address these questions.

We find support for the ``happiness is assortative'' hypothesis and evidence that these correlations can be detected up to three links away. Further, this finding does not appear to be based on users tweeting similar words (Fig. \ref{fig:shuffled_hamming}). Our correlation coefficients for reciprocal-reply networks constructed at the level of weeks are smaller than those obtained by Bollen et al.\ \cite{Bollen2011} for a reciprocal-follower network constructed by aggregating over a six month period. This difference is  likely a reflection of differences in methodologies, such as our more dynamic time scale (one-week periods vs. six month periods), our exclusion of central value happiness scores (i.e., stop words), and our use of the Spearman correlation coefficient. 

While this paper does not attempt to separate homophily and contagion, future work could use reciprocal-reply networks to investigate these effects. While reciprocal-reply networks are subject to errors caused by missing data (see above discussion of this issue) they may provide a valuable framework for studying contagion effects, given that they are based on a conservative and dynamic metric of what constitutes an interaction on Twitter. A network structure in which links are known to be active and valid provides an arena in which the diffusion of information and emotion may be properly studied.

\section*{Acknowledgments}
The authors acknowledge the Vermont Advanced Computing Core which is supported by NASA (NNX-08AO96G) at the University of Vermont for providing High Performance Computing resources that have contributed to the research results reported within this paper. CAB was supported by the UVM Complex Systems Center Fellowship Award, KDH was supported by VT NASA EPSCoR, and PSD was supported by NSF Career Award \# 0846668. CMD and PSD were also supported by a grant from the MITRE Corporation. 

\bibliographystyle{model1-num-names}

\bibliography{twitter_refs}
  \renewcommand{\thefigure}{A\arabic{equation}}    
  \renewcommand{\thetable}{A\arabic{equation}}    
  \setcounter{equation}{1}  % reset counter     

\section*{Appendix}
\onecolumn
\begin{figure}[htp!]{
\centering     
\includegraphics[width=.45\textwidth]{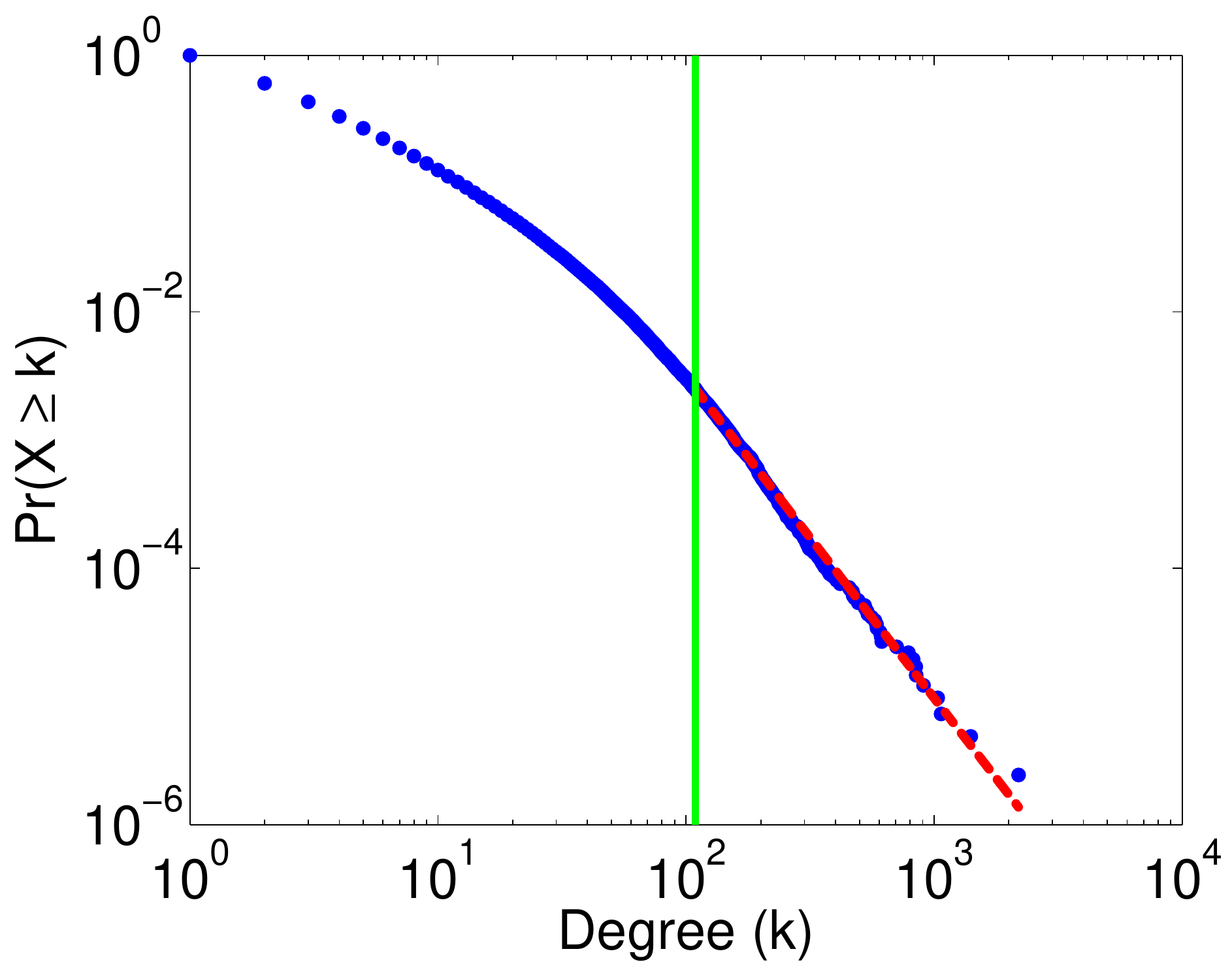}
\caption{Log-log plot of the complementary cumulative distribution function (CCDF) of the degree distribution for a sample month (January 2009) network is shown (blue), along with the best fitting power law model ($\alpha=3.50$ and $k_{\min}=109$) using the procedure of Clauset, Shalizi, and Newman \cite{clauset_powerlaw}. We test whether the empirical distribution is distinguishable from a power law using the Kolmogorov-Smirnov test and find no evidence against the null hypothesis ($D=1.82 \times 10^{-2} , p=0.35, n=495881$) data. This distribution may be fit by a power law.}
\label{fig:prk_month}}
     \end{figure}
\setcounter{equation}{2}
      \begin{figure}[htp!]
\centering 
       \includegraphics[width=.45\textwidth]{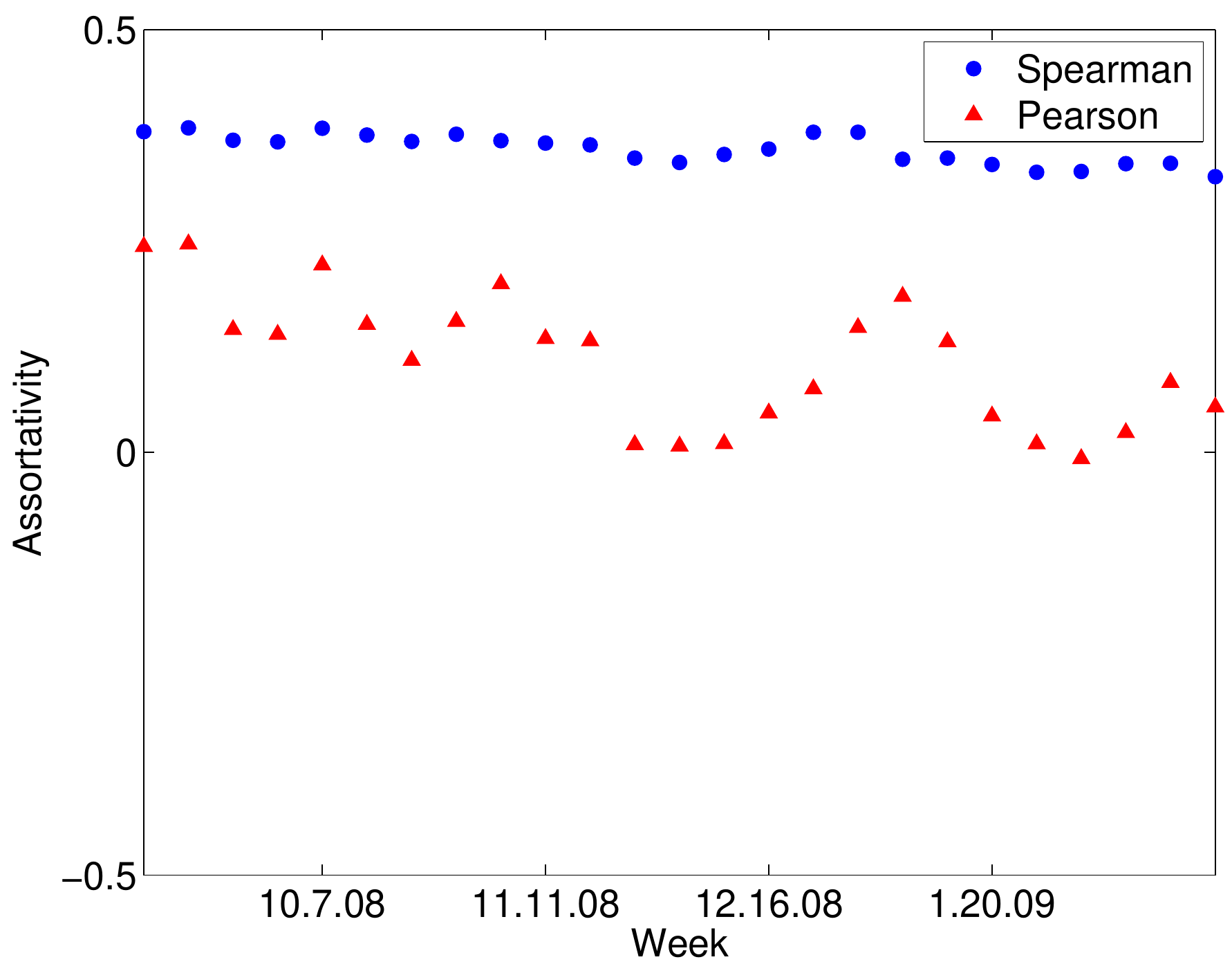} 
\caption{Spearman and Pearson correlation coefficients are used to measure degree assortativity. The Pearson correlation coefficient is more sensitive to extreme values. As a result, the Pearson correlation coefficient obscures the trend that the network is assortative with respect to the rank of node degrees. Given the nature of the degree distribution and the questions that we are asking, we use the Spearman correlation coefficient for our study.}
  \label{fig:spearvspearlastone}
     \end{figure} 

\setcounter{equation}{3}
\begin{figure}[htp]
\centering     
     \subfigure[Assortativity of happiness, Pearson's $r$]{
      \includegraphics[width=.70\textwidth, height=9cm]{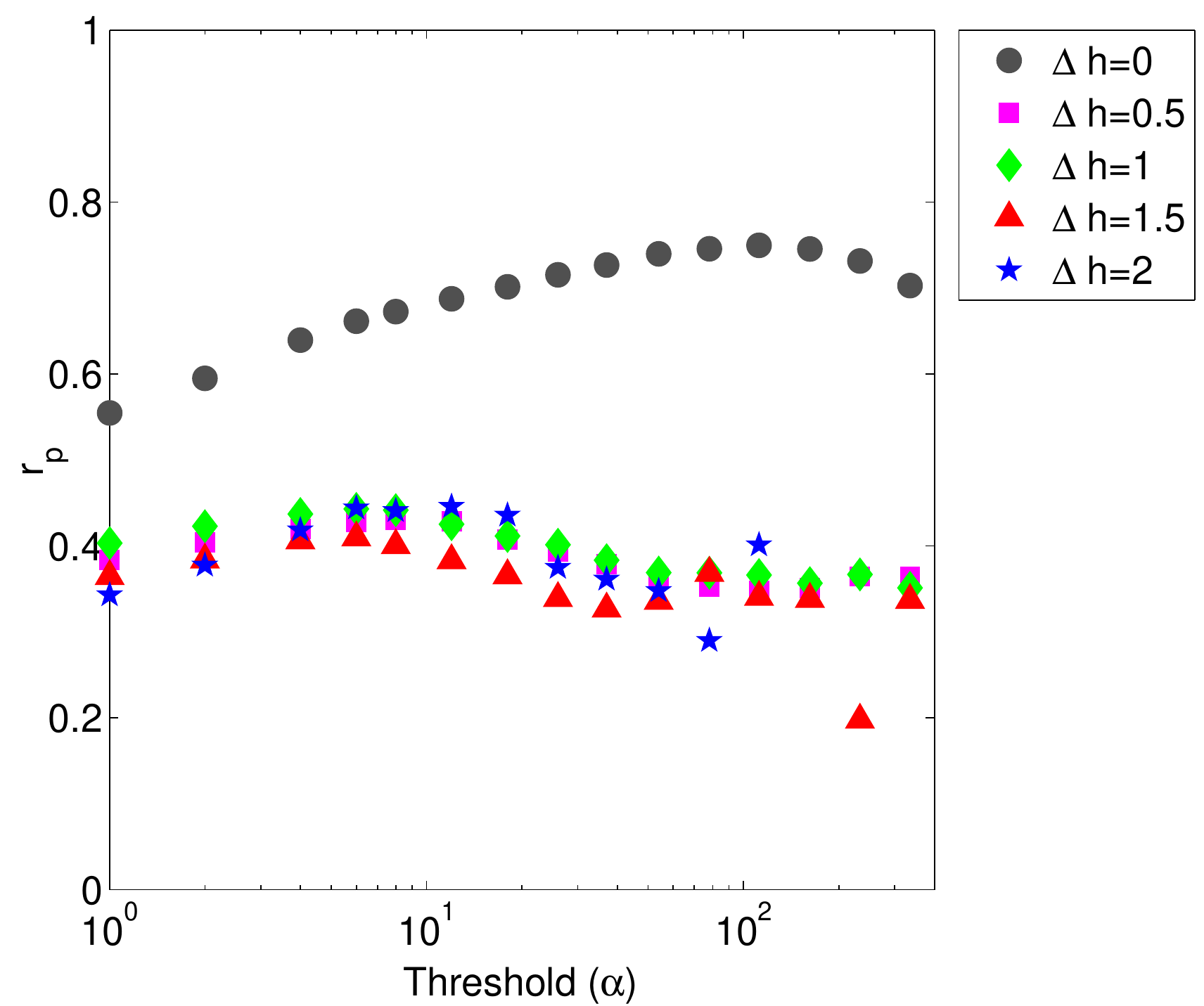}}\\
\subfigure[Assortativity of happiness, Spearman's $r$]{
            \includegraphics[width=.70\textwidth, height=9cm]{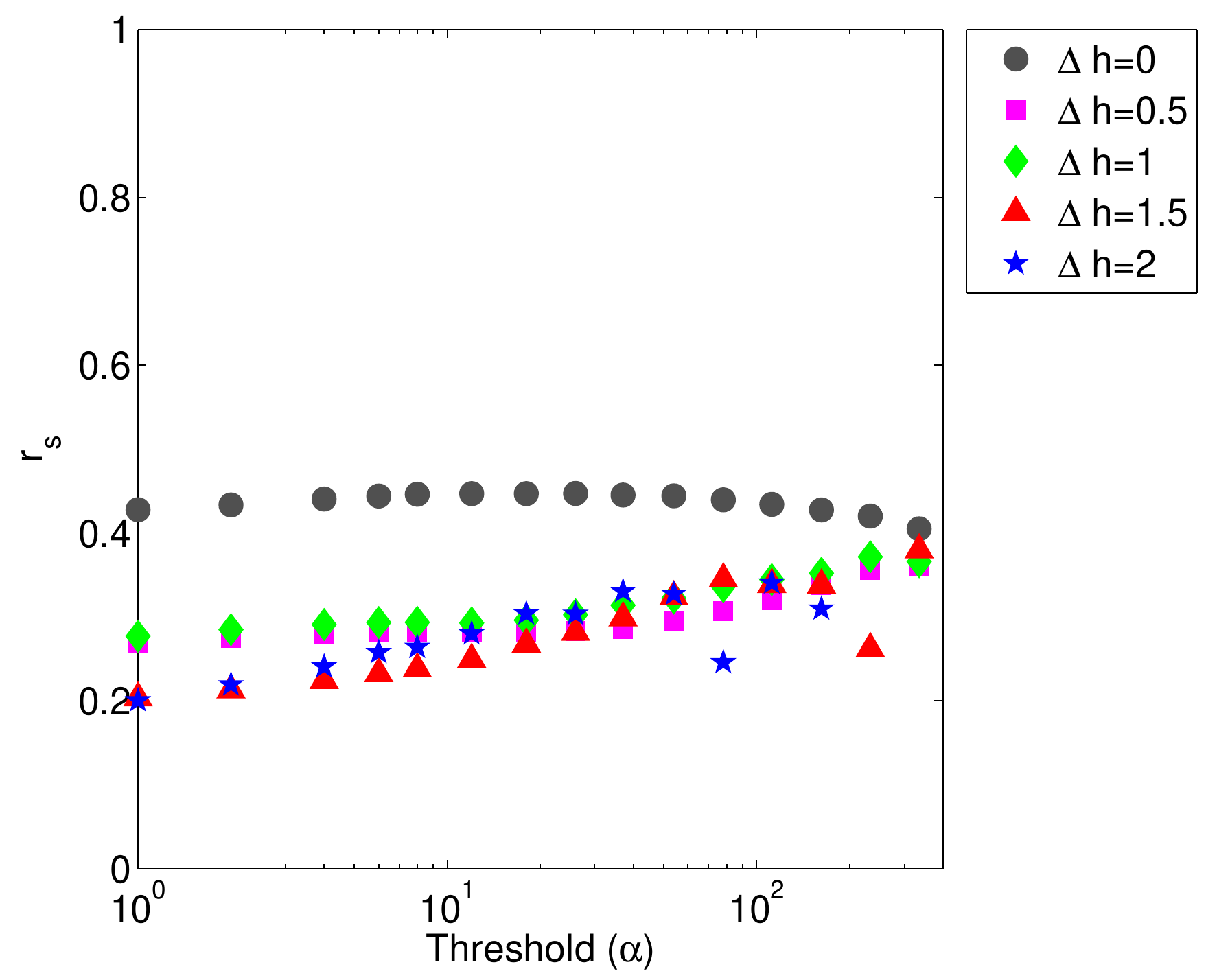}}\\
      \caption{Measured happiness assortativity as threshold for labMT word usage increases for a single week network. The Spearman correlation coefficient (right) exhibits less variability as compared to the Pearson correlation coefficient (left). Notice that when $\Delta h=0$, there is less variation due to the numerous words centered around the mean happiness score, regardless of the threshold, $\alpha$. }
            \label{fig:spearman_vs_pearson_dial_deltah}
\end{figure}

\onecolumn
\setcounter{equation}{4}
\begin{figure*}[htp!]
\centering     
        \includegraphics[width=\textwidth]{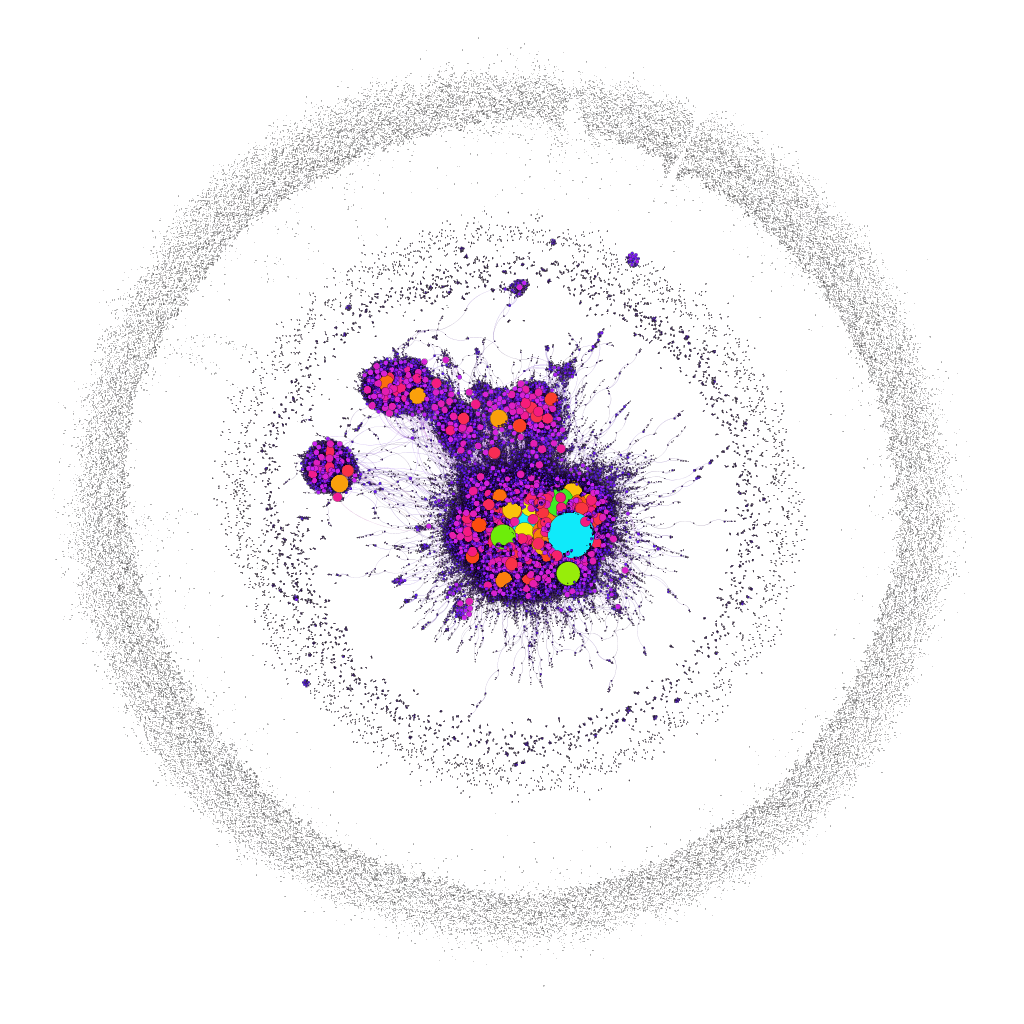}
\caption{A visualization of the reciprocal reply network for the week beginning September 9, 2008 (Week 1) is depicted. The size of a node is proportional to the degree, and colors further emphasize the degree detected by Gephi's implementation of the algorithm suggested by Blondel et al. \cite{forceatlas2}.}
      \label{fig:week1_white}
     \end{figure*} 
\setcounter{equation}{5}
\begin{figure*}[htp!]
\centering     
                      \subfigure{\includegraphics[width=\textwidth]{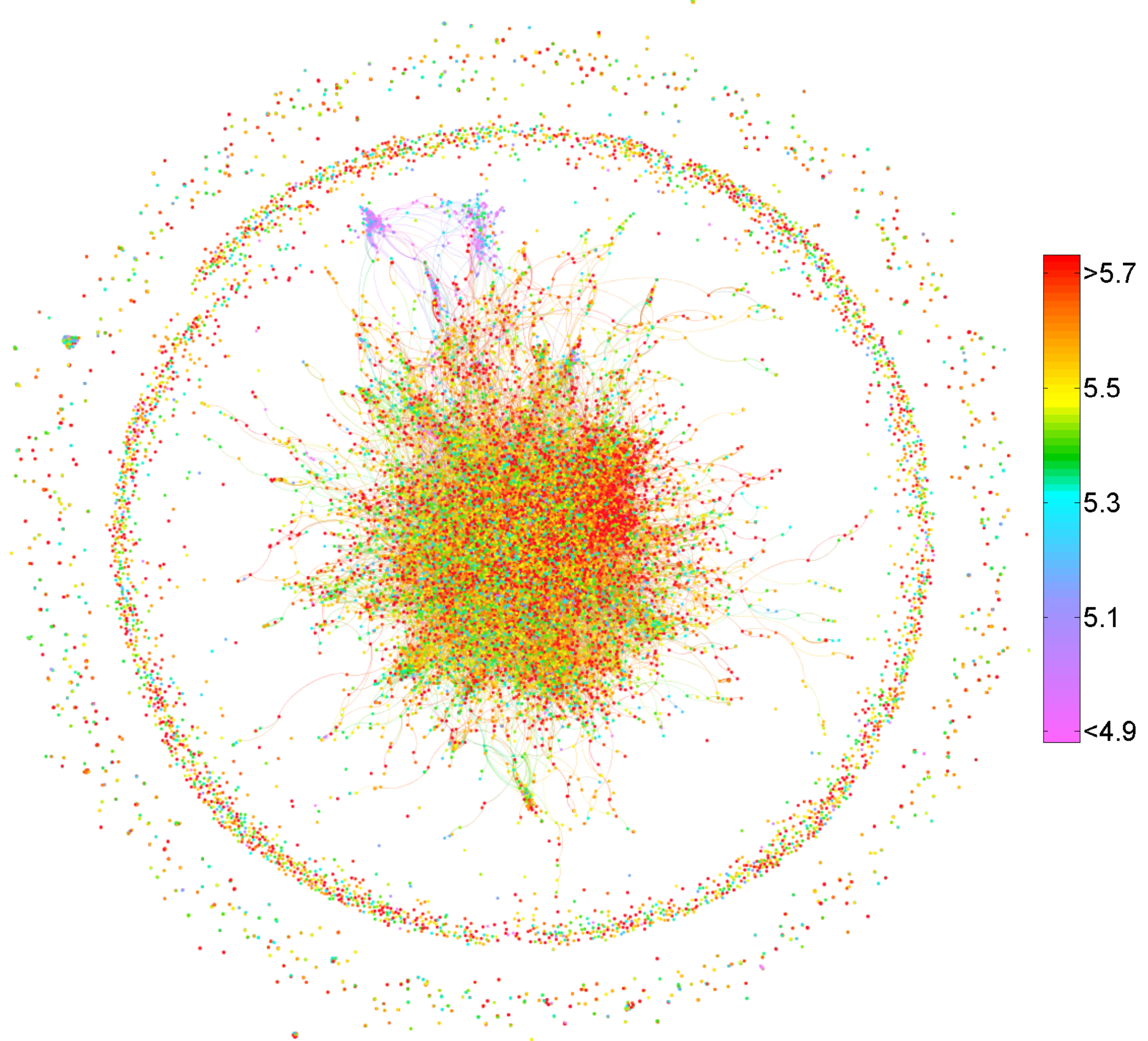}}                             
\caption{A visualization of the reciprocal reply network for the week beginning September 9, 2008 (Week 1). Colors represent happiness scores for nodes with greater than $\alpha = 50$ labMT words (57\% of all nodes in the week). The visualization was produced using Gephi \cite{gephi}. The algorithm employed by the software clusters nodes according to their connectivity. Collections of nodes with similar colors provide a visualization of the happiness is assortativity finding.}
\label{fig:pighate}
     \end{figure*}  

\setcounter{equation}{6}

\begin{figure*}[htp!]
\centering     
        \includegraphics[width=\textwidth]{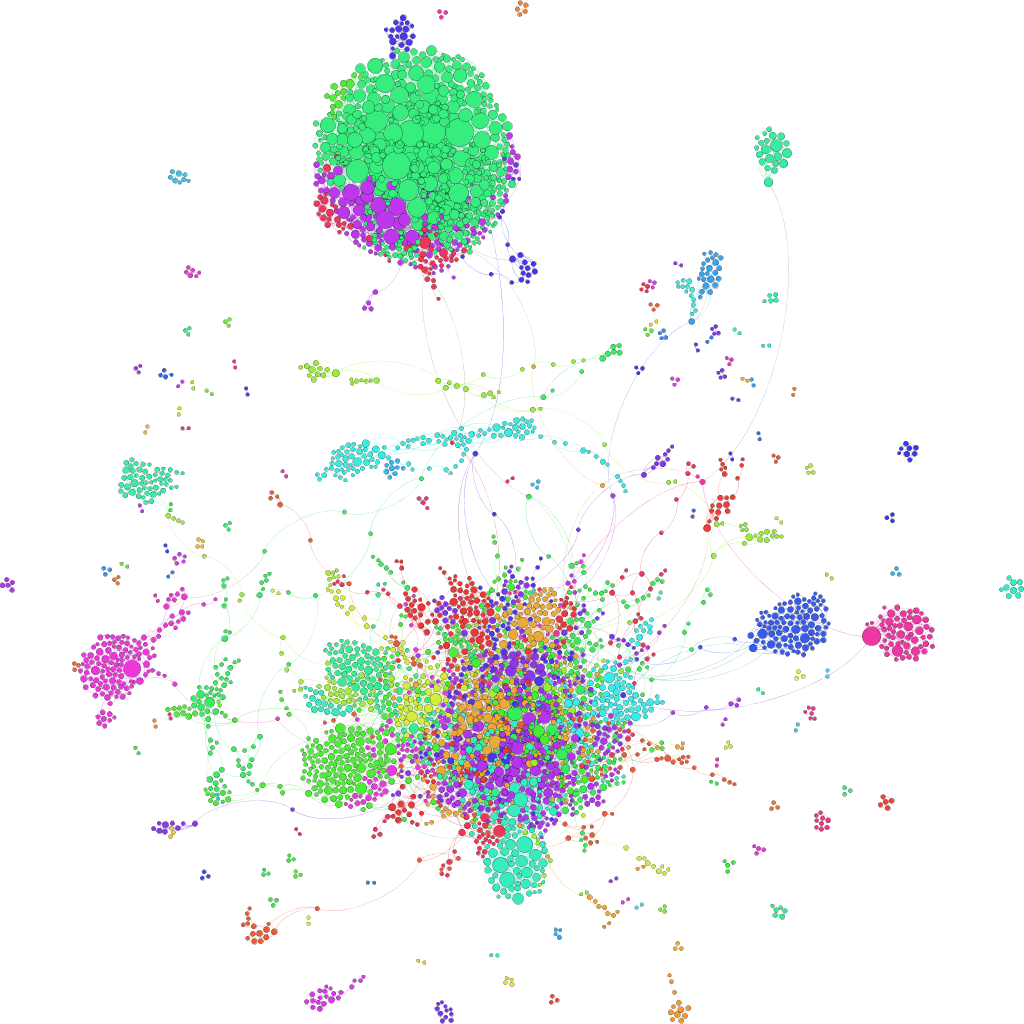}
\caption{A visualization of the reciprocal reply network for the day of October 28, 2008. The size of the nodes is proportional to the degree and colors indicate communities detected by Gephi's implementation of the community detection algorithm suggested by \cite{blondel2008}.}.
      \label{fig:day_modularity}
     \end{figure*} 
\setcounter{equation}{7}

\begin{figure}[htp!]{
\centering     
\includegraphics[width=.6\textwidth]{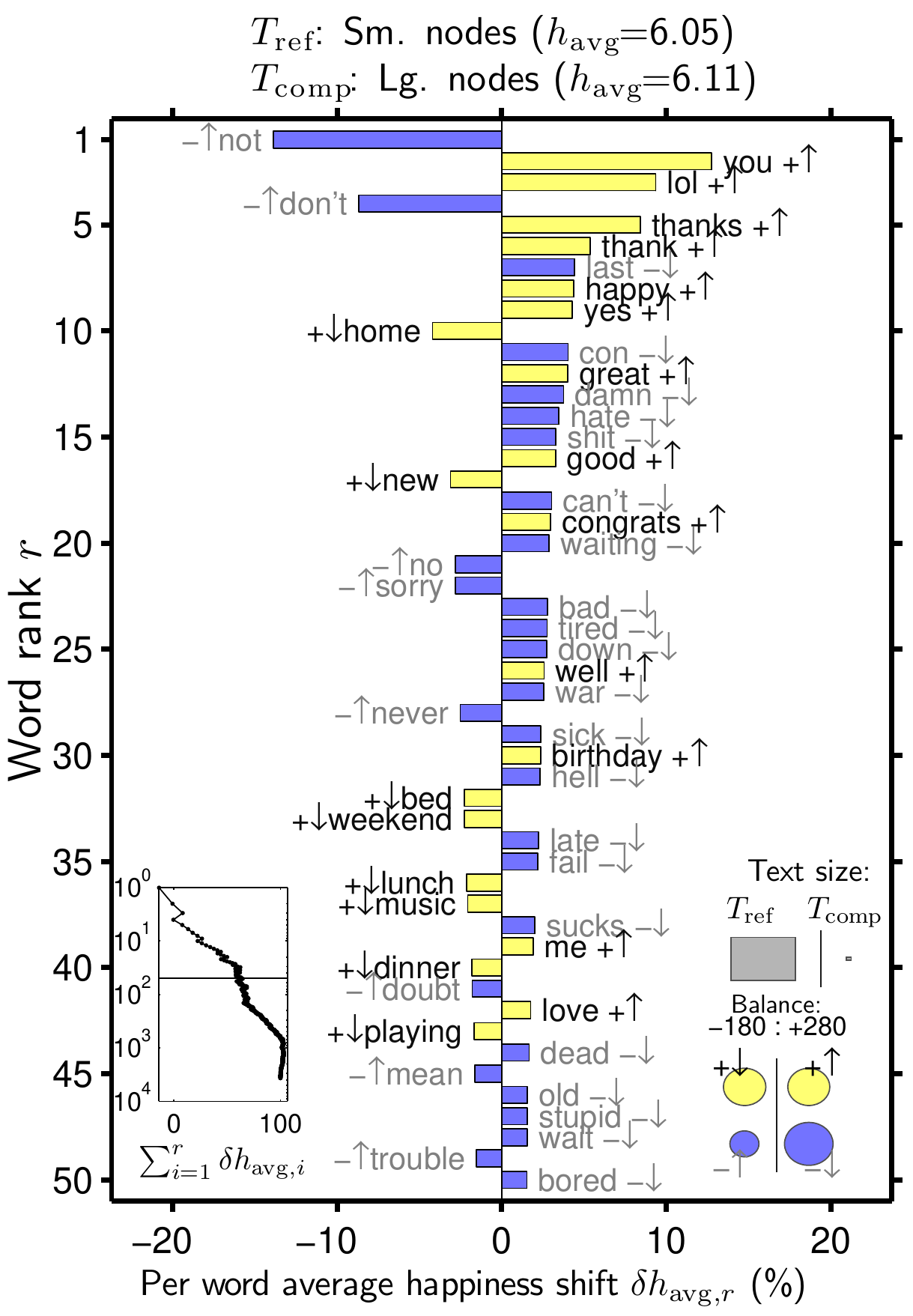}
\caption{The collection of words used by nodes with degree $k \geq 100$ ($T_{\text{comp}}$) is compared to words written by users with degree $k<100$ ($T_{\text{ref}}$). Words ``award'' and ``awards'' were excluded because their usage dominated the wordbag of one high degree node (Twitter's ShortyAwards). The word ``die'' was also excluded to eliminate the possibility of the hedonometer incorrectly being applied to German tweets. We note that removal of these words resulted in a negligible change to all values of $r_s$ reported in the paper. The horizontal bars on the right side of the plot represent words which raise the happiness score of $T_{\text{comp}}$. The symbols of $+/-$ and $\uparrow / \downarrow$ combine to convey whether a positive/negative word appears more/less frequently in the $T_{\text{comp}}$ as compared to the $T_{\text{ref}}$. Notice that an increase in the usage of positive words (e.g., ``you''), as well as a decrease in the use of a negative word (e.g., ``last'') will contribute to $T_{\text{comp}}$ having a higher happiness score. On the left hand side of the word shift plot are words which contribute to lowering the happiness score of $T_{\text{comp}}$. Such examples include an increase in the usage of negative words (e.g., ``not'') as well as a decrease in the usage of positive words (e.g., ``home''). The magnitude of the bars indicate the relative contribution of each word to these effects. In summary, we see that $T_{\text{comp}}$ has a higher happiness score than does $T_{\text{ref}}$. In the lower right, the relative text sizes are depicted as rectangles proportional to the number of words. The reference text, $T_{\text{ref}}$, has considerably more words in its collection than does $T_{\text{comp}}$. The circle plots depicted the relative amount of positive vs. negative words contained in $T_{\text{ref}}$ and $T_{\text{comp}}$. While both collections are similar in terms of positive word usage, the collection of words used by larger nodes contains fewer negative words and thus, this contributes to the slightly higher happiness score for this collection of words. The lower left inset shows the cumulative sum of individual word contributions as a function of $\log_{10} r$, where $r$ is the rank of the 3,686 labMT words. See \cite{Dodds2011} for the full details of the wordshift graph.} 
\label{fig:worshift}}
     \end{figure} 

     \setcounter{equation}{8}

           \begin{figure}[htp!]
\centering 
       \includegraphics[width=.8\textwidth]{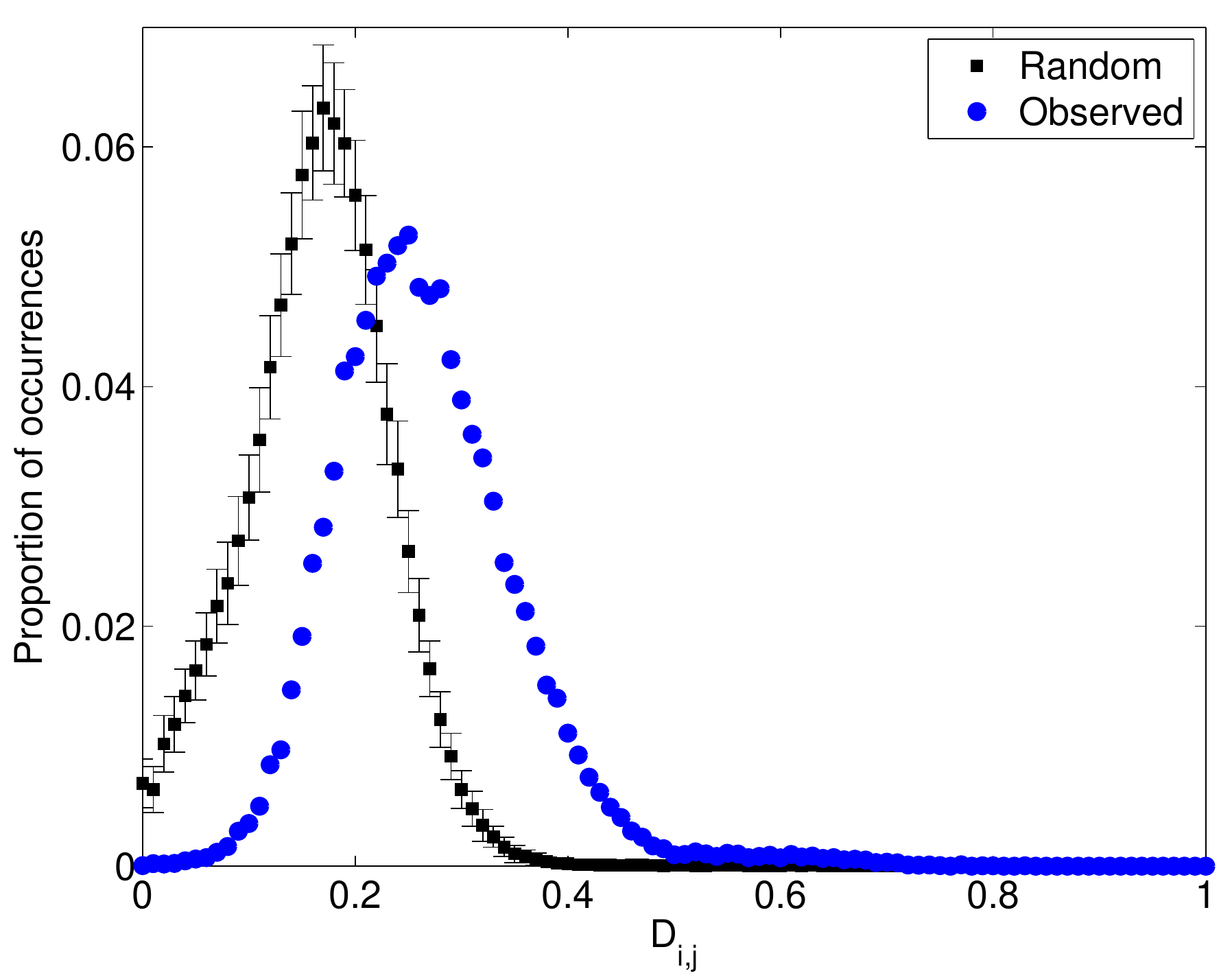} 
       
\caption{The similarity of word bags for pairs of users connected in a week reciprocal reply network is computed as follows: For users $i$ and $j$, we compute $D_{i,j}=1-\frac{1}{2}\sum^{3686}_{n=1} |f_{i,n}-f_{j,n}|$, where $f_{i,n}$ represents the normalized frequency of word usage of the $n$th labMT word by user $i$. The value of $D_{i,j}$ ranges from 0 (dissimilar word bags) to 1 (similar word bags). The proportion of occurrences of user-user pairs in the reciprocal reply network for a sample week (Sept. 16, 2008) having word similarity indices between 0 and 1 are shown (blue dots), with $\alpha=50$ and $\Delta h=1$. The majority of user-user similarity indices are less than 0.4, indicating that users and their nearest neighbors use dissimilar collections of words in their tweets. We then perform 100 random permutations of word vector assignments to users, while holding the network topology intact (black squares). The resulting distributions show that while users are using more similar words than would be expected by chance, this shift is small. The mean score for randomized user-user paired word collections is $\overline{D_{i,j}}=.167$. This value is not zero, since users are using a common language (English). The mean score for our observed network data is $\overline{D_{i,j}}=.267$, which is slightly higher than the randomized value due to conversations occurring between these users.} 
  \label{fig:shuffled_hamming}
     \end{figure} 
\setcounter{equation}{1}
\begin{table*}[ht!]
\label{table:A1}
\begin{center}
\tabcolsep=0.11cm
\begin{tabular}{rrrr|rrrrr|rrrrr}
\hline
Rank & Word & Frequency & Happiness & & Rank & Word & Frequency & Happiness & & Rank & Word & Frequency & Happiness  \\ 
     &  & $(\times 10^5)$ &         & &      &  & $(\times 10^5)$ &         & &      &  & $(\times 10^5)$ &         \\
\hline 
\hline
1 &           you & 103.55 & 6.24 & & 51 &          sure & 6.82 & 6.32 & & 101 &         music & 4.24 & 8.02     \\ 
 2 &            my & 94.91 & 6.16 & & 52 &          done & 6.81 & 6.54 & & 102 &         found & 4.23 & 6.54     \\ 
 3 &            me & 56.35 & 6.58 & & 53 &          show & 6.73 & 6.24 & & 103 &       doesn't & 4.23 & 3.62     \\ 
 4 &           not & 39.98 & 3.86 & & 54 &       awesome & 6.72 & 7.60 & & 104 &        online & 4.23 & 6.72     \\ 
 5 &            up & 36.04 & 6.14 & & 55 &         check & 6.51 & 6.10 & & 105 &         party & 4.20 & 7.58     \\ 
 6 &            no & 34.40 & 3.48 & & 56 &           bed & 6.42 & 7.18 & & 106 &          soon & 4.20 & 6.34     \\ 
 7 &           new & 34.03 & 6.82 & & 57 &         sleep & 6.33 & 7.16 & & 107 &      thinking & 4.15 & 6.28     \\ 
 8 &          like & 31.75 & 7.22 & & 58 &          cool & 6.32 & 7.20 & & 108 &          snow & 4.14 & 6.32     \\ 
 9 &           all & 30.71 & 6.22 & & 59 &          live & 6.29 & 6.84 & & 109 &          give & 4.13 & 6.54     \\ 
10 &          good & 30.20 & 7.20 & & 60 &           big & 6.28 & 6.22 & & 110 &         movie & 4.12 & 6.84     \\ 
11 &          will & 23.58 & 6.02 & & 61 &          free & 6.18 & 7.96 & & 111 &            ha & 4.09 & 6.00     \\ 
12 &            we & 22.59 & 6.38 & & 62 &          life & 6.17 & 7.32 & & 112 &         sorry & 4.08 & 3.66     \\ 
13 &           day & 21.80 & 6.24 & & 63 &           old & 6.07 & 3.98 & & 113 &          real & 4.06 & 6.78     \\ 
14 &          know & 19.45 & 6.10 & & 64 &        didn't & 6.04 & 4.00 & & 114 &          kids & 3.98 & 7.38     \\ 
15 &          more & 19.32 & 6.24 & & 65 &          find & 6.00 & 6.00 & & 115 &         phone & 3.91 & 6.44     \\ 
16 &         don't & 18.29 & 3.70 & & 66 &           die & 6.00 & 1.74 & & 116 &            tv & 3.91 & 6.70     \\ 
17 &         today & 18.24 & 6.22 & & 67 &         video & 5.99 & 6.48 & & 117 &          stop & 3.89 & 3.90     \\ 
18 &          love & 17.66 & 8.42 & & 68 &         house & 5.99 & 6.34 & & 118 &          play & 3.88 & 7.26     \\ 
19 &         think & 17.45 & 6.20 & & 69 &     christmas & 5.89 & 7.96 & & 119 &       waiting & 3.88 & 3.68     \\ 
20 &           see & 15.28 & 6.06 & & 70 &       playing & 5.77 & 7.14 & & 120 &         lunch & 3.81 & 7.42     \\ 
21 &         great & 14.60 & 7.88 & & 71 &         world & 5.76 & 6.52 & & 121 &          food & 3.79 & 7.44     \\ 
22 &           lol & 13.35 & 6.84 & & 72 &          game & 5.54 & 6.92 & & 122 &       reading & 3.76 & 6.78     \\ 
23 &        thanks & 13.09 & 7.40 & & 73 &           wow & 5.54 & 7.46 & & 123 &           god & 3.74 & 7.28     \\ 
24 &          home & 13.05 & 7.14 & & 74 &         ready & 5.53 & 6.58 & & 124 &           top & 3.65 & 6.76     \\ 
25 &        people & 12.71 & 6.16 & & 75 &        iphone & 5.53 & 6.54 & & 125 &           buy & 3.60 & 6.28     \\ 
26 &         night & 12.70 & 6.22 & & 76 &     listening & 5.41 & 6.28 & & 126 &          book & 3.56 & 7.24     \\ 
27 &          blog & 12.26 & 6.02 & & 77 &        pretty & 5.40 & 7.32 & & 127 &           car & 3.56 & 6.72     \\ 
28 &          last & 11.89 & 3.74 & & 78 &        always & 5.39 & 6.48 & & 128 &          idea & 3.52 & 7.06     \\ 
29 &          well & 11.70 & 6.68 & & 79 &          help & 5.27 & 6.08 & & 129 &        friend & 3.51 & 7.66     \\ 
30 &          make & 11.27 & 6.00 & & 80 &          read & 5.07 & 6.52 & & 130 &        family & 3.51 & 7.72     \\ 
31 &         right & 11.04 & 6.54 & & 81 &        google & 5.05 & 7.20 & & 131 &           yay & 3.47 & 6.10     \\ 
32 &         can't & 10.93 & 3.42 & & 82 &      everyone & 5.03 & 6.12 & & 132 &          glad & 3.47 & 7.48     \\ 
33 &       morning & 10.38 & 6.56 & & 83 &          most & 4.95 & 6.22 & & 133 &         least & 3.46 & 4.00     \\ 
34 &          very & 10.10 & 6.12 & & 84 &          wait & 4.88 & 3.74 & & 134 &       nothing & 3.44 & 3.90     \\ 
35 &         first & 9.69 & 6.82 & & 85 &         start & 4.87 & 6.10 & & 135 &          late & 3.43 & 3.46     \\ 
36 &           our & 9.26 & 6.08 & & 86 &        please & 4.79 & 6.36 & & 136 &      internet & 3.39 & 7.48     \\ 
37 &        better & 8.89 & 7.00 & & 87 &           con & 4.78 & 3.70 & & 137 &       amazing & 3.38 & 7.66     \\ 
38 &            us & 8.82 & 6.26 & & 88 &           try & 4.77 & 6.02 & & 138 &          mean & 3.38 & 3.68     \\ 
39 &       tonight & 8.79 & 6.14 & & 89 &       thought & 4.69 & 6.38 & & 139 &        myself & 3.37 & 6.30     \\ 
40 &          down & 8.73 & 3.66 & & 90 &        school & 4.66 & 6.26 & & 140 &      facebook & 3.34 & 6.08     \\ 
41 &         happy & 8.40 & 8.30 & & 91 &         thank & 4.64 & 7.40 & & 141 &         funny & 3.32 & 7.92     \\ 
42 &      tomorrow & 7.88 & 6.18 & & 92 &       weekend & 4.56 & 8.00 & & 142 &         tired & 3.29 & 3.34     \\ 
43 &          nice & 7.80 & 7.38 & & 93 &           hey & 4.48 & 6.06 & & 143 &          talk & 3.29 & 6.06     \\ 
44 &          best & 7.61 & 7.18 & & 94 &          wish & 4.44 & 6.92 & & 144 &          damn & 3.26 & 2.98     \\ 
45 &           she & 7.57 & 6.18 & & 95 &          hate & 4.42 & 2.34 & & 145 &   interesting & 3.26 & 7.52     \\ 
46 &           yes & 7.42 & 6.74 & & 96 &          haha & 4.41 & 7.64 & & 146 &           own & 3.24 & 6.16     \\ 
47 &           fun & 7.37 & 7.96 & & 97 &       friends & 4.41 & 7.92 & & 147 &        friday & 3.23 & 6.88     \\ 
48 &          hope & 7.34 & 7.38 & & 98 &        making & 4.40 & 6.24 & & 148 &          open & 3.18 & 6.10     \\ 
49 &           bad & 6.98 & 2.64 & & 99 &        dinner & 4.27 & 7.40 & & 149 &          lost & 3.16 & 2.76     \\ 
50 &         never & 6.92 & 3.34 & & 100 &        coffee & 4.27 & 7.18 & & 150 &          guys & 3.16 & 6.22     \\ 
\hline
\end{tabular}
\end{center}
\caption{The top 150 most frequently occurring words from the labMT list in our Sept 2008 through Feb 2009 data set, where stop words ($4<h_{\text{avg}}<6$) have been removed.}
\end{table*}

\newpage

\setcounter{equation}{2}
\begin{table*}[ht]
\label{table:A2}
\begin{center}
\tabcolsep=0.11cm
\begin{tabular}{rrrr|rrrrr|rrrrr}
\hline
Rank & Word & Frequency & Happiness & & Rank & Word & Frequency & Happiness & & Rank & Word & Frequency & Happiness  \\ 
     &  & $(\times 10^5)$ &         & &      &  & $(\times 10^5)$ &         & &      &  & $(\times 10^5)$ &         \\
\hline 
\hline
 1 &           the & 295.60 & 4.98 & & 51 &            an & 22.73 & 4.84 & & 101 &            el & 11.76 & 4.80     \\ 
 2 &            to & 249.91 & 4.98 & & 52 &            we & 22.59 & 6.38 & & 102 &          well & 11.70 & 6.68     \\ 
 3 &             i & 221.28 & 5.92 & & 53 &          some & 22.32 & 5.02 & & 103 &            oh & 11.69 & 4.84     \\ 
 4 &             a & 218.13 & 5.24 & & 54 &           que & 22.26 & 4.64 & & 104 &           who & 11.64 & 5.06     \\ 
 5 &           and & 135.23 & 5.22 & & 55 &           day & 21.80 & 6.24 & & 105 &        should & 11.48 & 5.24     \\ 
 6 &            is & 127.94 & 5.18 & & 56 &           how & 21.64 & 4.68 & & 106 &          over & 11.34 & 4.82     \\ 
 7 &            in & 122.94 & 5.50 & & 57 &         going & 20.64 & 5.42 & & 107 &          make & 11.27 & 6.00     \\ 
 8 &            of & 121.79 & 4.94 & & 58 &            am & 20.60 & 5.38 & & 108 &          then & 11.15 & 5.34     \\ 
 9 &           for & 114.41 & 5.22 & & 59 &            go & 20.03 & 5.54 & & 109 &         right & 11.04 & 6.54     \\ 
10 &           you & 103.55 & 6.24 & & 60 &           has & 19.68 & 5.18 & & 110 &         can't & 10.93 & 3.42     \\ 
11 &            on & 96.97 & 5.56 & & 61 &            or & 19.55 & 4.98 & & 111 &           way & 10.84 & 5.24     \\ 
12 &            my & 94.91 & 6.16 & & 62 &          know & 19.45 & 6.10 & & 112 &          only & 10.72 & 4.92     \\ 
13 &            it & 91.09 & 5.02 & & 63 &          more & 19.32 & 6.24 & & 113 &       getting & 10.63 & 5.68     \\ 
14 &          that & 69.81 & 4.94 & & 64 &            la & 18.77 & 5.00 & & 114 &           his & 10.56 & 5.56     \\ 
15 &            at & 58.51 & 4.90 & & 65 &         don't & 18.29 & 3.70 & & 115 &       morning & 10.38 & 6.56     \\ 
16 &          with & 56.42 & 5.72 & & 66 &         today & 18.24 & 6.22 & & 116 &          very & 10.10 & 6.12     \\ 
17 &            me & 56.35 & 6.58 & & 67 &           too & 18.15 & 5.22 & & 117 &         after & 9.82 & 5.08     \\ 
18 &          just & 50.25 & 5.76 & & 68 &          they & 18.09 & 5.62 & & 118 &      watching & 9.76 & 5.84     \\ 
19 &          have & 49.86 & 5.82 & & 69 &          work & 17.95 & 5.24 & & 119 &           her & 9.73 & 5.84     \\ 
20 &            be & 46.10 & 5.68 & & 70 &           got & 17.91 & 5.60 & & 120 &          them & 9.71 & 4.92     \\ 
21 &          this & 45.75 & 5.06 & & 71 &          love & 17.66 & 8.42 & & 121 &         first & 9.69 & 6.82     \\ 
22 &            de & 44.38 & 4.82 & & 72 &         think & 17.45 & 6.20 & & 122 &             e & 9.66 & 4.72     \\ 
23 &            so & 40.93 & 5.08 & & 73 &          back & 17.37 & 5.18 & & 123 &        that's & 9.55 & 5.28     \\ 
24 &           not & 39.98 & 3.86 & & 74 &       twitter & 17.18 & 5.46 & & 124 &            rt & 9.52 & 4.88     \\ 
25 &           i'm & 39.89 & 5.74 & & 75 &          when & 16.84 & 4.96 & & 125 &             y & 9.47 & 4.48     \\ 
26 &           are & 39.03 & 5.16 & & 76 &         there & 16.39 & 5.10 & & 126 &          than & 9.42 & 4.74     \\ 
27 &           but & 37.78 & 4.24 & & 77 &           had & 15.30 & 4.74 & & 127 &           its & 9.36 & 4.96     \\ 
28 &           was & 37.74 & 4.60 & & 78 &           see & 15.28 & 6.06 & & 128 &           our & 9.26 & 6.08     \\ 
29 &            up & 36.04 & 6.14 & & 79 &            en & 14.97 & 4.84 & & 129 &        better & 8.89 & 7.00     \\ 
30 &           out & 35.20 & 4.62 & & 80 &        really & 14.93 & 5.84 & & 130 &            us & 8.82 & 6.26     \\ 
31 &           now & 35.12 & 5.90 & & 81 &           off & 14.89 & 4.02 & & 131 &       tonight & 8.79 & 6.14     \\ 
32 &            no & 34.40 & 3.48 & & 82 &         great & 14.60 & 7.88 & & 132 &          down & 8.73 & 3.66     \\ 
33 &           new & 34.03 & 6.82 & & 83 &          need & 14.45 & 4.84 & & 133 &          i've & 8.59 & 5.58     \\ 
34 &            do & 33.96 & 5.76 & & 84 &            he & 14.34 & 5.42 & & 134 &             u & 8.40 & 5.52     \\ 
35 &          from & 33.78 & 5.18 & & 85 &         still & 13.74 & 5.14 & & 135 &         happy & 8.40 & 8.30     \\ 
36 &          like & 31.75 & 7.22 & & 86 &          been & 13.43 & 5.04 & & 136 &         again & 8.34 & 5.42     \\ 
37 &          your & 31.43 & 5.60 & & 87 &           lol & 13.35 & 6.84 & & 137 &         could & 8.34 & 5.52     \\ 
38 &           all & 30.71 & 6.22 & & 88 &         would & 13.15 & 5.38 & & 138 &            un & 8.15 & 4.64     \\ 
39 &          good & 30.20 & 7.20 & & 89 &        thanks & 13.09 & 7.40 & & 139 &          into & 8.08 & 5.04     \\ 
40 &           get & 30.04 & 5.92 & & 90 &          home & 13.05 & 7.14 & & 140 &          i'll & 8.05 & 5.38     \\ 
41 &          what & 29.46 & 4.80 & & 91 &          want & 12.81 & 5.70 & & 141 &           man & 7.99 & 5.90     \\ 
42 &         about & 28.97 & 5.16 & & 92 &        people & 12.71 & 6.16 & & 142 &      tomorrow & 7.88 & 6.18     \\ 
43 &          it's & 27.14 & 4.88 & & 93 &         night & 12.70 & 6.22 & & 143 &          nice & 7.80 & 7.38     \\ 
44 &            if & 25.21 & 4.66 & & 94 &          here & 12.28 & 5.48 & & 144 &           any & 7.70 & 5.22     \\ 
45 &            by & 24.66 & 4.98 & & 95 &             o & 12.26 & 4.96 & & 145 &          take & 7.63 & 5.18     \\ 
46 &            as & 24.50 & 5.22 & & 96 &          blog & 12.26 & 6.02 & & 146 &          best & 7.61 & 7.18     \\ 
47 &          time & 24.19 & 5.74 & & 97 &           why & 12.10 & 4.98 & & 147 &           she & 7.57 & 6.18     \\ 
48 &           one & 23.73 & 5.40 & & 98 &          much & 11.92 & 5.74 & & 148 &          even & 7.42 & 5.58     \\ 
49 &          will & 23.58 & 6.02 & & 99 &          last & 11.89 & 3.74 & & 149 &           yes & 7.42 & 6.74     \\ 
50 &           can & 23.57 & 5.62 & & 100 &           did & 11.84 & 5.58 & & 150 &        little & 7.38 & 4.60     \\ 
 \hline
\end{tabular}
\end{center}
\caption{The top 150 most frequently occurring words from the labMT word list in our Sept 2008 through Feb 2009 data set including stop words.}
\end{table*}

\newpage
 \setcounter{equation}{3}  

\begin{table*}[ht!]
\label{table:n s}
\begin{center}
\begin{tabular}{rrrrrrrrr}
\hline 
Week & Start date &$N$  & $<k>$ & $k_{\max}$ & $C_G$ & Assort & \# Comp. & $S$  \\[1ex]
\hline\hline                            
 1 & 09.09.08 & 95647 & 2.99 & 261 & 0.10 & 0.24 & 10364 & 0.71   \\ 
 2 & 09.16.08 & 99236 & 2.95 & 313 & 0.10 & 0.24 & 11062 & 0.71   \\ 
 3 & 09.23.08 & 99694 & 2.90 & 369 & 0.09 & 0.13 & 11457 & 0.70   \\ 
 4 & 09.30.08 & 100228 & 2.87 & 338 & 0.09 & 0.13 & 11752 & 0.69   \\ 
 5 & 10.07.08 & 78296 & 2.60 & 241 & 0.09 & 0.21 & 11140 & 0.63   \\ 
 6 & 10.14.08 & 122644 & 3.20 & 394 & 0.09 & 0.14 & 12221 & 0.74   \\ 
 7 & 10.21.08 & 130027 & 3.30 & 559 & 0.08 & 0.09 & 12420 & 0.75   \\ 
 8 & 10.28.08 & 144036 & 3.56 & 492 & 0.08 & 0.14 & 12319 & 0.78   \\ 
 9 & 11.04.08 & 145346 & 3.54 & 330 & 0.08 & 0.19 & 12597 & 0.78   \\ 
10 & 11.11.08 & 136534 & 3.35 & 441 & 0.08 & 0.12 & 12972 & 0.76   \\ 
11 & 11.18.08 & 153486 & 3.46 & 444 & 0.08 & 0.13 & 13594 & 0.77   \\ 
12 & 11.25.08 & 155753 & 3.46 & 1244 & 0.06 & 0.00 & 14122 & 0.77   \\ 
13 & 12.02.08 & 165156 & 3.44 & 1245 & 0.06 & 0.01 & 14496 & 0.78   \\ 
14 & 12.09.08 & 162445 & 3.33 & 1456 & 0.05 & 0.01 & 15342 & 0.76   \\ 
15 & 12.16.08 & 148154 & 3.12 & 730 & 0.06 & 0.04 & 15645 & 0.73   \\ 
16 & 12.23.08 & 140871 & 3.22 & 575 & 0.07 & 0.07 & 15216 & 0.72   \\ 
17 & 12.30.08 & 143015 & 3.30 & 519 & 0.07 & 0.15 & 15272 & 0.73   \\ 
18 & 01.06.09 & 170597 & 3.19 & 253 & 0.07 & 0.18 & 17234 & 0.74   \\ 
19 & 01.13.09 & 188429 & 3.29 & 477 & 0.07 & 0.13 & 18403 & 0.75   \\ 
20 & 01.20.09 & 196038 & 3.16 & 680 & 0.06 & 0.04 & 19927 & 0.74   \\ 
21 & 01.27.09 & 203852 & 3.04 & 973 & 0.05 & 0.01 & 21537 & 0.73   \\ 
22 & 02.03.09 & 212513 & 2.92 & 1718 & 0.04 & -0.01 & 24387 & 0.71   \\ 
23 & 02.10.09 & 213936 & 2.83 & 828 & 0.06 & 0.02 & 25854 & 0.70   \\ 
24 & 02.17.09 & 215172 & 2.65 & 437 & 0.06 & 0.07 & 28742 & 0.67   \\ 
25 & 02.24.09 & 170180 & 2.27 & 320 & 0.06 & 0.04 & 28388 & 0.58   \\ 
\hline
\end{tabular}
\end{center}
\caption{Network statistics for reciprocal-reply networks by week. As Twitter popularity grows, so does the number of users ($N$) in the observed reciprocal-reply network. The average degree ($<k>$), degree assortativity, the number of nodes in the giant component (\# Comp.), and the proportion of nodes in the giant component ($S$) remain fairly constant, whereas the maximum degree ($k_{\max}$) shows a great deal of variability from month to month. Clustering ($C_G$) shows a slight decrease over the course of this period.}
\end{table*}
 \setcounter{equation}{4}  

\begin{table*}[ht!]
\label{table:data set}
\begin{center}
\begin{tabular}{lllllll}
\hline
\hline                                                  
Week & Start date	& \# Obsvd.	Msgs. & \# Total Msgs.	& \% Obsvd. & \# Replies	& \%  Replies  \\
& & $\times 10^6$ & $\times 10^6$ &  $\left( \frac{\# Obsvd.}{\# Total} \times 100 \right)$  & $\times 10^6$ &  $\left( \frac{\# Replies}{\# Obsvd.}  \times 100\right)$  \\[1ex]
\hline\hline                            
 1 & 09.09.08 & 3.14 & 7.26 & 43.2 &  0.88 & 28.1 \\ 
 2 & 09.16.08 & 3.36 & 8.31 & 40.4 &  0.90 & 26.9 \\ 
 3 & 09.23.08 & 3.43 & 8.89 & 38.6 &  0.90 & 26.2 \\ 
 4 & 09.30.08 & 3.33 & 9.06 & 36.8 &  0.89 & 26.6 \\ 
 5 & 10.07.08 & 2.33 & 9.38 & 24.8 &  0.64 & 27.5 \\ 
 6 & 10.14.08 & 4.39 & 9.87 & 44.4 &  1.24 & 28.3 \\ 
 7 & 10.21.08 & 4.70 & 10.01 & 47.0 &  1.35 & 28.8 \\
 8 & 10.28.08 & 5.74 & 10.34 & 55.5 &  1.64 & 28.5 \\ 
 9 & 11.04.08 & 5.58 & 11.14 & 50.1 &  1.63 & 29.3 \\ 
10 & 11.11.08 & 4.70 & 9.88 & 47.6 &  1.42 & 30.2 \\ 
11 & 11.18.08 & 5.48 & 11.34 & 48.3 &  1.67 & 30.5 \\ 
12 & 11.25.08 & 5.71 & 11.47 & 49.8 &  1.73 & 30.2 \\ 
13 & 12.02.08 & 5.54 & 12.85 & 43.1 &  1.80 & 32.4 \\ 
14 & 12.09.08 & 5.41 & 13.54 & 39.9 &  1.72 & 31.7 \\ 
15 & 12.16.08 & 4.57 & 12.72 & 35.9 &  1.45 & 31.8 \\ 
16 & 12.23.08 & 4.80 & 11.62 & 41.3 &  1.46 & 30.5 \\ 
17 & 12.30.08 & 4.61 & 13.48 & 34.2 &  1.50 & 32.5 \\ 
18 & 01.06.09 & 5.16 & 16.11 & 32.0 &  1.72 & 33.3 \\ 
19 & 01.13.09 & 5.73 & 17.33 & 33.1 &  1.97 & 34.4 \\ 
20 & 01.20.09 & 5.82 & 18.87 & 30.9 &  1.98 & 34.1 \\ 
21 & 01.27.09 & 5.75 & 20.79 & 27.6 &  1.98 & 34.5 \\ 
22 & 02.03.09 & 5.78 & 22.42 & 25.8 &  2.01 & 34.8 \\ 
23 & 02.10.09 & 5.66 & 23.39 & 24.2 &  1.99 & 35.1 \\ 
24 & 02.17.09 & 5.43 & 25.71 & 21.1 &  1.91 & 35.1 \\ 
25 & 02.24.09 & 3.80 & 20.75 & 18.3 &  1.34 & 35.1 \\ 

\hline
\end{tabular}
\end{center}
\caption{The number of ``observed'' messages in our database comprise a fraction of the total number of Twitter message made during period of this study (September 2008 through February 2009). While our feed from the Twitter API remains fairly constant, the total \# of tweets grows, thus reducing the \% of all tweets observed in our database. We calculate the total $\#$ of messages as the difference between the last message id and the first message id that we observe for a given month. This provides a reasonable estimation of the number of tweets made per month as message ids were assigned (by Twitter) sequentially during the time period of this study. We also report the number observed messages that are replies to specific messages and the percentage of our observed messages which constitute replies. }
\end{table*}

\newpage

\end{document}